\documentclass[
 preprint,
 aps, pra,
 amsmath,amssymb,
 11pt,
 final,
tightenlines,
 twoside,
 twocolumn,
 nofloats,
 nobibnotes,
 nofootinbib,
 superscriptaddress,
 noshowpacs,
 showkeys,
 showkeywords,
 centertags]{revtex4-2}

\usepackage[T2A]{fontenc}
\usepackage[utf8x]{inputenc}
\usepackage[english]{babel}
\usepackage{graphicx}
\usepackage{dcolumn}
\usepackage{bm}
\usepackage{longtable}

\input{maik.rty}

\setcitestyle{authoryear,round}
\def\saoname{Special Astrophysical Observatory,  Russian Academy of Sciences,
              Nizhnii Arkhyz, 369167 Russia}

\def\squareforqed{\hbox{\rlap{$\sqcap$}$\sqcup$}}

\def\sq{\ifmmode\squareforqed\else{\unskip\nobreak\hfil
\penalty50\hskip1em\null\nobreak\hfil\squareforqed
\parfillskip=0pt\finalhyphendemerits=0\endgraf}\fi}

\def\degr{\hbox{$^\circ$}}

\def\utw{\smash{\rlap{\lower5pt\hbox{$\sim$}}}}

\def\udtw{\smash{\rlap{\lower6pt\hbox{$\approx$}}}}

\def\fd{\hbox{$\,.\!\!^{\rm d}$}}

\def\fdg{\hbox{$\,.\!\!^\circ$}}

\def\farcs{\hbox{$\,.\!\!^{\prime\prime}$}}

\def\diameter{{\ifmmode\mathchoice
{\ooalign{\hfil\hbox{$\displaystyle/$}\hfil\crcr
{\hbox{$\displaystyle\mathchar"20D$}}}}
{\ooalign{\hfil\hbox{$\textstyle/$}\hfil\crcr
{\hbox{$\textstyle\mathchar"20D$}}}}
{\ooalign{\hfil\hbox{$\scriptstyle/$}\hfil\crcr
{\hbox{$\scriptstyle\mathchar"20D$}}}}
{\ooalign{\hfil\hbox{$\scriptscriptstyle/$}\hfil\crcr
{\hbox{$\scriptscriptstyle\mathchar"20D$}}}}
\else{\ooalign{\hfil/\hfil\crcr\mathhexbox20D}}%
\fi}}

\newcommand{\aap}{Astron. and Astrophys. }

\newcommand{\aaps}{Astron. and Astrophys. Suppl. }

\newcommand{\aj}{Astron.~J. }
\renewcommand{\apj}{Astrophys.~J. }
\newcommand{\apjs}{Astrophys.~J. Suppl. }

\newcommand{\araa}{Annual Rev. Astron. Astrophys. }

\newcommand{\mnras}{Monthly Notices Royal Astron. Soc. }

\newcommand{\apjl}{Astrophys.~J.}

\begin{document}

\selectlanguage{english}
\keywords{stars: magnetic field---stars: chemically peculiar}




\title{Magnetic Fields of CP Stars in the Orion\,OB1 Association. \\IV. Stars of Subgroup 1b}

\author{\firstname{I. I.}~\surname{Romanyuk}} \email{roman@sao.ru} \affiliation{\saoname}

\author{\firstname{E. A.}~\surname{Semenko}} \affiliation{\saoname}
\affiliation{National Astronomical Research Institute of Thailand, Chiangmai, 50180 Thailand}
\author{\firstname{A. V.}~\surname{Moiseeva}} \affiliation{\saoname}

\author{\firstname{I. A.}~\surname{Yakunin}} \affiliation{\saoname} \affiliation{Saint Petersburg State University, Saint Petersburg, 199034 Russia}

\author{\firstname{D. O.}~\surname{Kudryavtsev}} \affiliation{\saoname}


\begin{abstract}
The paper presents magnetic field measurements for 15~chemically
peculiar (CP) stars of subgroup~1b in the Orion\,OB1 association.
We have found that the proportion of stars with strong magnetic
fields among these 15~CP stars is almost twice as large as in
subgroup~1a. Along with this, the age of subgroup~1b is estimated
as 2~Myr, and the age of subgroup~1a is in the order of 10~Myr.
The average root-mean-square magnetic field $\langle B_{e}
\rangle({\rm all})$ for stars in subgroup~1b is 2.3~times higher
than that for stars in subgroup~1a. The conclusions obtained fall
within the concept of the fossil origin of large-scale magnetic
fields in B and A stars, but the rate of field weakening with age
appears anomalously high. We present our results as an important
observational test for calibrating the theory of stellar magnetic
field formation and evolution.
\end{abstract}

\maketitle

\section{INTRODUCTION}

The development of the theory of stellar magnetic field formation
and evolution requires reliable observational tests which can show
reliability of theoretical conclusions. As opposed to the Sun,
where the formation and development of the field is observed
directly on the surface, in the case of magnetic chemically
peculiar stars everything is much more complicated. Here the
observation of the field structure in detail is impossible,
therefore we can only compare  the results of our observations
with magnetohydrodynamic calculations, taking into account many
parameters, which often are not exactly quantifiable.

The question about a dependence between the magnetic field
strength and the age of a star  is one of the most important in
the research of magnetic fields in Ap/Bp~stars. Without mentioning
specific works, we can note that researchers generally agree that
there is a trend of weakening of the field and simplification of
its geometry with age. However, these conclusions often appear to
be insufficiently reliable due to the difficulties in determining
the age of single stars, hence the intense interest in magnetic
stars in clusters and associations.

\begin{table*}[ht!!!]
\caption{Summary data on CP stars of subgroup~1b of the Orion
OB1\, association, included in the research program} \label{tab1}
\medskip
\begin{tabular}{l|c|c|c|c|c|c|c}
\hline
HD number &Peculiarity &$l$, degr. &$b$, degr. &$\pi$, mas &$d$, pc &$m_V$, mag. &$A_V$, mag. \\
\hline
36046  &   He-wk     & 203.74 & $-18.57$ & 2.91 & 343 & 8.06 & 0.15 \\
36313  &   He-wk, Si & 203.77 & $-18.05$ & 3.17 & 315 & 8.20 & 0.12 \\
36485  &   He-r      & 203.84 & $-17.73$ & 2.57 & 389 & 6.85 & 0.12 \\
36526  &   He-wk, Si & 205.08 & $-18.31$ & 2.44 & 410 & 8.29 & 0.18 \\
36668  &   He-wk, Si & 203.18 & $-16.98$ & 2.36 & 424 & 8.07 & 0.01 \\
36955  &   CrEu      & 205.25 & $-17.58$ & 2.29 & 437 & 9.58 & --   \\
37140  &   He-wk     & 204.39 & $-16.79$ & 2.43 & 412 & 8.56 & 0.69 \\
37149  &   He-wk     & 205.62 & $-17.42$ & 2.38 & 420 & 8.02 & 0.05 \\
37235  &   He-wk     & 204.84 & $-16.84$ & 2.51 & 398 & 8.13 & 0.06 \\
37321  &   He-wk     & 205.58 & $-17.04$ & 1.56 & 640 & 7.09 & 0.17 \\
37333  &   Si        & 206.54 & $-17.50$ & 2.85 & 350 & 8.51 & 0.22 \\
37479  &   He-r      & 206.81 & $-17.32$ & 2.28 & 438 & 6.34 & 0.25 \\
37525  &   He-wk     & 206.89 & $-17.29$ & 2.29 & 436 & 8.06 & 0.17 \\
37633  &   EuSi      & 207.01 & $-17.14$ & 2.40 & 417 & 9.01 & 0.44 \\
37776  &   He-r      & 206.07 & $-16.34$ & 2.28 & 438 & 6.99 & 0.28 \\
290665 &   CrEuSr    & 204.74 & $-17.29$ & 2.48 & 403 & 9.44 & 0.19 \\
\hline
\end{tabular}
\end{table*}

In our research program, we decided to measure magnetic fields of
all chemically peculiar stars in the Orion\,OB1 association with
reliably determined ages of its individual subgroups. This article
continues the series of our papers \citep{1,2,3,4,5}
for complex investigation of magnetic chemically peculiar stars in
Orion. The objectives of the program were presented in \citet{1},
the observation and data analysis method was described in
\citet{2,3,4}.

Since our first publication \citep{1}, a lot of new information
has been accumulated in the world about the properties of the
Orion\,OB1 association. The most important for our study are the
results of the GAIA mission, which made it possible to establish
the exact distances to individual stars of the association. These
data in many respects confirmed the correctness of our estimates
of the distances to those association members for which the
parallax was either absent or its value diverged from other
observed characteristics.

The final aim of our research program is to analyze possible
evolutionary effects observed in Ori~OB1 chemically peculiar
stars, particularly answer the questions of whether the occurrence
of peculiar stars relative to normal stars depends on age and how
the occurrence of magnetic CP stars varies with age. By now, we
have completed the study of the oldest subgroup~1a, the average
age of the stars of which is about 10~Myr \citep{5}. In this next
paper of the series, we present magnetic field measurements for CP
stars of subgroup~1b of the association. The presentation of the
results is similar to that in the paper of~\citet{5}.

Subgroup~1b of the Orion\,OB1 association forms, almost entirely,
the Orion belt asterism. All the previous numerous studies show
that this region of the constellation belongs to the youngest in
the association. The average age of stars in it is estimated as
2~Myr.  Accuracy and reliability of these estimates are much
higher than those in the case of age determination using
spectroscopic data and evolutionary tracks.

\section{MAGNETIC FIELD MEASUREMENTS}
\subsection{Selection of Stars to Observe}

According to the method proposed in \citet{1}, we selected for
observations sixteen CP stars in subgroup~1b of the Orion\,OB1
association. Table~\ref{tab1} gives information about individual
stars: the object number in the HD catalog, the peculiarity type
according to the catalog of \citet{6}, the galactic coordinates
$l$ and $b$, the parallax $\pi$ obtained by Gaia, the distance $d$
to the star in parsecs, the magnitude $m_{V}$ in the $V$ band, and
the total interstellar or circumstellar extinction $A_{V}$ towards
the star \citep{1}.

Table~\ref{tab1} shows that the selected stars are located rather
densely in the space within the galactic coordinates $l=[203\fdg2;
206\fdg9]$ and $b=[-16\fdg3; -18\fdg6]$ at distances from 315 to
440~pc. The only exception would be the binary star HD\,37321, the
parallax of which could be determined erroneously. Thus, the
region occupied by the subgroup stars has a size of about
$30\times15$~pc in the perspective plane and no more than 100 pc
in depth along the line of sight.

\subsection{Measurements}

Between 2015 and 2019, for each of the selected stars we obtained
at least five spectra with a Zeeman analyzer~\citep{8} at the Main
Stellar Spectrograph~(MSS)~\citep{7} of the 6-m telescope (BTA).
The observation and data reduction techniques are completely
similar to those used earlier (see \citet{5}).

Table~\ref{tab2} shows the magnetic field measurements. The
columns of the table present: the number of the star in the HD
catalog; the Julian date of an observation; the signal-to-noise
ratio $S/N$; the longitudinal magnetic field in gauss,  obtained
using the modified Babcock method ($B_{e}(z)$,) the regression
method ($B_{e}(r)$), and by the H$\beta$ hydrogen line
($B_{e}(h)$); the corresponding root-mean-square errors $\sigma$.
The magnetic-field measurement method is described in detail in
the previous papers \citep{2,3,4,5}. For sixteen stars, we
obtained a total of 110 magnetic field measurements, at least five
for each object.

For statistical studies we use the root-mean-square magnetic field $\langle B_{e}\rangle$, its error $\sigma$, and the$\chi^2/n$ value characterizing the reliability of field detection against the background of measurement errors (see formulas~(\ref{for1})--(\ref{for3}) adopted from \citep{27}). We consider a star to be magnetic if $\chi^2/n > 5$.\\[-20pt]
\begin{equation}
    \label{for1}
    \langle B_{e}\rangle =\left(\dfrac{1}{n} \sum_{i=1}^n B_{{e}i}^2\right)^{1/2}
    \end{equation}\\[-25pt]
\begin{equation}
    \label{for2}
    \langle \sigma\rangle = \left(\dfrac{1}{n} \sum_{i=1}^n \sigma_{i}^2\right)^{1/2}
\end{equation}\\[-25pt]
\begin{equation}
    \label{for3}
    \chi^2/n = \dfrac{1}{n} \sum_{i=1}^n \left(\dfrac{B_{{e}i}}{\sigma_{i}}\right)^{2}
\end{equation}\\[-20pt]

Below, we comment on the measurements for each star. In addition,
we supplement the information about the stars with our
measurements of radial $V_{R}$ and rotation $v_{e}\sin i$
velocities, and also present the corresponding values from
astronomical databases.

\onecolumngrid
\begin{longtable}{l|c|c|r@{$\,\pm\,$}l|r@{$\,\pm\,$}l|c}
\caption{Magnetic field measurements for CP stars in subgroup~1b of the Orion\,OB1 association}\label{tab2}\\
\hline
Star    & JD, (2450000+) & $S/N$ & \multicolumn{2}{c|}{$B_{e}(z)\pm\sigma$, G} & \multicolumn{2}{c|}{$B_{e}(r)\pm\sigma$, G} & $B_{e}(h)$, G \\
\hline
\endfirsthead
\caption{(Continued) }\\
\hline
Star    & JD, (2450000+) & $S/N$ & \multicolumn{2}{c|}{$B_{e}(z)\pm\sigma$, G} & \multicolumn{2}{c|}{$B_{e}(r)\pm\sigma$, G} & $B_{e}(h)$, G \\
\hline
\endhead

\hline
\endfoot

\hline
\endlastfoot

HD\,36046  &  $6640.320$  & $220$ & $-750$   & $1410$       & $-100$                  & $90$   & --     \\
           &  $7740.416$  & $200$ & $910$    & $790$        & $180$                   & $460$  & $100$    \\
           &  $8125.437$  & $250$ & $-50$    & $1200$       & $-150$                  & $260$  & $-400$   \\
           &  $8151.199$  & $210$ & $-1690$  & $610$        & $-70$                   & $150$  & $-1200$  \\
           &  $8446.449$  & $260$ & $-1170$  & $1570$       & $-30$                   & $110$  & $-250$   \\
\hline
HD\,36313  &  $5554.321$  & $400$ & $120$    & $120$        & $560$                   & $180$  & $-1800$  \\
           &  $5842.500$  & $350$ & $160$    & $160$        & $480$                   & $190$  & $1600$    \\
           &  $5843.542$  & $350$ & $40$     & $130$        & $250$                   & $270$  & $1100$    \\
           &  $6995.325$  & $250$ & $-160$   & $400$        & $280$                   & $210$  & $2000$    \\
           &  $7288.512$  & $350$ & $-480$   & $370$        & $340$                   & $150$  & --    \\
           &  $7288.528$  & $400$ & $-300$   & $300$        & $-70$                   & $190$  & $600$    \\
           &  $7289.492$  & $300$ & $280$    & $160$        & $-40$                   & $120$  & $-500$  \\
           &  $7290.483$  & $300$ & $-60$    & $290$        & $20$                    & $180$  & $-1500$  \\
           &  $8830.358$  & $220$ & $340$    & $150$        & $280$                   & $130$  & $1900$    \\
           &  $8830.452$  & $240$ & $110$    & $290$        & $-200$                  & $210$  & $-2700$  \\
           &  $8830.507$  & $180$ & $20$     & $190$        & $-200$                  & $150$  & $-2000$  \\
           &  $8834.479$  & $140$ & $760$    & $170$        & $370$                   & $180$  & $700$    \\
\hline
HD\,36485  &  $5553.247$  & $330$ & $-2350$  & $250$        & $-2310$                 & $120$  & $-2100$  \\
           &  $5553.480$  & $330$ & $-2330$  & $220$        & $-2210$                 & $180$  & $-1900$  \\
           &  $5554.263$  & $300$ & $-2400$  & $210$        & $-2270$                 & $120$  & $-2300$  \\
           &  $5554.481$  & $300$ & $-2700$  & $230$        & \multicolumn{2}{c|}{--} & $-3100$  \\
           &  $5555.253$  & $330$ & $-2830$  & $260$        & $-2470$                 & $160$  & $-2600$  \\
           &  $5555.486$  & $270$ & $-2830$  & $320$        & $-2370$                 & $120$  & $-2400$  \\
           &  $5582.279$  & $370$ & $-2310$  & $240$        & $-2350$                 & $120$  & $-2500$  \\
           &  $5583.280$  & $320$ & $-3030$  & $260$        & $-2250$                 & $120$  & $-2600$  \\
           &  $5873.540$  & $300$ & $-3440$  & $320$        & $-2220$                 & $140$  & $-2700$  \\
           &  $5962.381$  & $390$ & $-2860$  & $320$        & $-2160$                 & $50$   & --     \\
           &  $5963.427$  & $320$ & $-2670$  & $210$        & $-2070$                 & $40$   & --     \\
\hline
HD\,36526  &  $5553.335$  & $250$ & $2730$   & $320$        & $2180$                  & $170$  & $3200$   \\
           &  $5842.532$  & $280$ & $1500$   & $400$        & $-290$                  & $210$  & $-2700$ \\
           &  $5963.292$  & $280$ & \multicolumn{2}{c|}{--} & $2790$                  & $50$   & --   \\
           &  $6589.530$  & $320$ & $2150$   & $220$        & $1970$                  & $130$  & $5700$   \\
           &  $7289.570$  & $200$ & $2730$   & $380$        & $1120$                  & $80$   & $3200$   \\
           &  $7290.525$  & $210$ & $4600$   & $600$        & $750$                   & $90$   & $1000$   \\
\hline
HD\,36668  &  $5582.396$  & $240$ & $-1040$  & $250$        & $-540$                  & $140$  & $-1350$  \\
           &  $5583.310$  & $310$ & $-1540$  & $220$        & $-1430$                 & $150$  & $-3300$  \\
           &  $5842.475$  & $300$ & $-1170$  & $350$        & $-1140$                 & $50$   & $-3400$  \\
           &  $5962.323$  & $300$ & $ 2170$  & $180$        & $1140$                  & $50$   & --      \\
           &  $5963.272$  & $300$ & $-1010$  & $780$        & $-920$                  & $60$   & --      \\
           &  $7288.565$  & $200$ & $ 2060$  & $350$        & $1030$                  & $80$   & $1700$  \\
           &  $7289.554$  & $240$ & $-3370$  & $650$        & $-500$                  & $140$  & $-1900$  \\
           &  $7290.513$  & $230$ & $ 1420$  & $430$        & $450$                   & $110$  & $5500$  \\
\hline
HD\,36955  &  $3273.529$  & $160$ & $-820$   & $190$        & \multicolumn{2}{c|}{--}          & --   \\
           &  $3274.512$  & $180$ & $-410$   & $200$        & \multicolumn{2}{c|}{--}          & --   \\
           &  $3275.510$  & $140$ & $-1300$  & $380$        & \multicolumn{2}{c|}{--}          & --   \\
           &  $4015.546$  & $280$ & $-480$   & $210$        & \multicolumn{2}{c|}{--}          & $50$   \\
           &  $6233.412$  & $110$ & $-750$   & $110$        & $-760$                  & $60$   & $-2100$ \\
           &  $8447.419$  & $170$ & $-970$   & $105$        & $-660$                  & $110$  & $-700$  \\
\hline
HD\,37140  &  $5555.297$  & $310$ & $-590$   & $90$         & $-350$                  & $90$   & --    \\
           &  $5962.400$  & $230$ & $220$    & $210$        & $ 220$                  & $50$   & --    \\
           &  $5963.440$  & $200$ & $140$    & $210$        & $-210$                  & $60$   & --    \\
           &  $8125.504$  & $120$ & $-900$   & $170$        & $-460$                  & $120$  & $-600$ \\
           &  $8151.227$  & $200$ & $-720$   & $220$        & $-80$                   & $140$  & $-100$ \\
           &  $8447.484$  & $150$ & $ 560$   & $290$        & $50$                    & $130$  & $-400$ \\
\hline
HD\,37149  &  $6643.344$  & $200$ & $5$      & $1200$       & $-320$                  & $120$  & --     \\
           &  $8008.545$  & $190$ & \multicolumn{2}{c|}{--} & $-180$                  & $190$  & $0$     \\
           &  $8446.475$  & $200$ & $-700$   & $1800$       & $-160$                  & $170$  & $800$    \\
           &  $8512.408$  & $240$ & $-3100$  & $3100$       & $170$                   & $190$  & $-400$ \\
           &  $8799.384$  & $250$ & $-1800$  & $2400$       & $150$                   & $180$  & $200$   \\
\hline
HD\,37235  &  $6643.321$  & $200$ & \multicolumn{2}{c|}{--} & $260$                   & $120$  & $-300$  \\
           &  $8126.173$  & $200$ & $9100$   & $4500$       & $170$                   & $170$  & $-900$  \\
           &  $8151.252$  & $180$ & $-700$   & $3800$       & $190$                   & $130$  & $200$  \\
           &  $8153.219$  & $190$ & $1370$   & $1330$       & $60$                    & $210$  & $-500$  \\
           &  $8447.509$  & $220$ & $1800$   & $2200$       & $380$                   & $160$  & $400$  \\
           &  $8550.243$  & $160$ & $90$     & $660$        & $180$                   & $210$  & $-800$  \\
\hline
HD\,37321  &  $6643.366$  & $280$ & $-590$   & $470$        & $-580$                  & $260$  & --  \\
           &  $7825.209$  & $200$ & $380$    & $650$        & $-160$                  & $210$  & $-400$  \\
           &  $8116.420$  & $260$ & $-250$   & $280$        & $10$                    & $140$  & --  \\
           &  $8153.247$  & $270$ & $-590$   & $460$        & $-180$                  & $200$  & $-200$  \\
           &  $8446.498$  & $400$ & $80$     & $210$        & $-150$                  & $180$  & $200$   \\
\hline
HD\,37333  &  $7762.470$  & $120$ & $50$     & $310$        & $-150$                  & $110$  & $-600$  \\
           &  $7823.188$  & $150$ & $560$    & $310$        & $70$                    & $110$  & $1300$  \\
           &  $8116.445$  & $140$ & $-1290$  & $270$        & $-440$                  & $130$  & $800$   \\
           &  $8446.525$  & $230$ & $-1030$  & $150$        & $-670$                  & $110$  & $-1200$  \\
           &  $8512.215$  & $160$ & $-890$   & $190$        & $-430$                  & $160$  & $-1700$  \\
           &  $8512.240$  & $140$ & $-850$   & $200$        & $-530$                  & $100$  & $-700$  \\
\hline
HD\,37479  &  $5555.324$  & $350$ & $-1050$  & $1080$       & $140$                   & $330$  & $-200$  \\
           &  $5582.343$  & $310$ & $4350$   & $540$        & $1630$                  & $270$  & $4800$   \\
           &  $5583.350$  & $280$ & $-3400$  & $1150$       & $-1860$                 & $480$  & $-4300$  \\
           &  $5963.347$  & $400$ & $2320$   & $360$        & $830$                   & $70$   & --   \\
\hline
HD\,37525  &  $5963.360$  & $280$ & $2390$   & $2770$       & $-20$                   & $90$   & --  \\
           &  $5555.337$  & $270$ & $670$    & $1670$       & $20$                    & $290$  & --  \\
           &  $7764.361$  & $260$ & $-780$   & $1080$       & $270$                   & $250$  & $-300$  \\
           &  $8446.553$  & $250$ & $620$    & $930$        & $-20$                   & $100$  & $1500$  \\
           &  $8514.409$  & $100$ & $-1460$  & $2130$       & $-100$                  & $140$  & $-1700$  \\
\hline
HD\,37633  &  $6643.421$  & $200$ & $400$    & $60$         & $320$                   & $80$   & $-400$  \\
           &  $7740.496$  & $500$ & $180$    & $170$        & $100$                   & $110$  & -- \\
           &  $8006.568$  & $140$ & $194$    & $190$        & $160$                   & $150$  & $-100$  \\
           &  $8126.455$  & $130$ & $810$    & $100$        & $660$                   & $80$   & $400$   \\
           &  $8447.365$  & $180$ & $740$    & $80$         & $460$                   & $63$   & $400$   \\
           &  $8758.511$  & $160$ & $440$    & $70$         & $300$                   & $80$   & $-400$  \\
\hline
HD\,37776  &  $8777.441$  & $320$ & $15700$  & $3900$       & $-140$                  & $155$  & $6600$    \\
           &  $8777.553$  & $370$ & $-2900$  & $2200$       & $-460$                  & $140$  & $-3100$  \\
           &  $8778.460$  & $350$ & $6400$   & $1800$       & $70$                    & $200$  & $7000$    \\
           &  $8778.578$  & $400$ & $-10100$ & $2900$       & $-580$                  & $160$  & $1500$   \\
           &  $8799.396$  & $310$ & $-2200$  & $2200$       & $210$                   & $180$  & $7000$   \\
           &  $8801.555$  & $290$ & $-800$   & $4000$       & $150$                   & $160$  & $7800$   \\
           &  $8805.369$  & $350$ & $-7500$  & $1000$       & $-1140$                 & $120$  & $-5300$ \\
           &  $8805.559$  & $350$ & $1300$   & $3000$       & $-60$                   & $190$  & $12300$   \\
           &  $8830.315$  & $300$ & $-3800$  & $2600$       & $-200$                  & $150$  & $10700$   \\
           &  $8830.478$  & $360$ & $-6700$  & $600$        & $-1240$                 & $110$  & $-4800$ \\
\hline
HD\,290665 &  $8007.527$  & $130$ & $3910$   & $120$        & $3200$                  & $40$   & $2800$   \\
           &  $8151.336$  & $180$ & $830$    & $90$         & $570$                   & $80$   & $1300$   \\
           &  $8447.442$  & $150$ & $1050$   & $100$        & $840$                   & $50$   & --  \\
           &  $8448.392$  & $150$ & $-3700$  & $140$        & $-2670$                 & $50$   & $-2500$ \\
           &  $8534.209$  & $190$ & $3450$   & $110$        & $2960$                  & $50$   & $3000$   \\
           &  $8535.178$  & $ 90$ & $1068$   & $120$        & $960$                   & $60$   & $400$   \\
           &  $8550.287$  & $130$ & $3210$   & $120$        & $2660$                  & $40$   & $2700$   \\
           &  $8551.250$  & $170$ & $-2990$  & $130$        & $-2240$                 & $40$   & $-1800$  \\
           &  $8579.179$  & $100$ & $-2800$  & $150$        & $-1720$                 & $60$   & $-300$  \\
           &  $8581.176$  & $150$ & $3410$   & $130$        & $2810$                  & $50$   & $1600$   \\
\end{longtable}
 \twocolumngrid

\subsection{Comments on Individual Stars}

\subsubsection{HD\,36046 = BD\,$-00\degr964$ \\= Renson\,9290 = Brown\,007b}

In the catalogs of variable stars of the Orion\,OB1
association, HD\,36046 is not presented. In the star spectrum in
the region of 4400--4970~\AA, there are several lines broadened by
rotation (\mbox {$v_{e}\sin i = 100$}~km s$^{-1}$). In the SIMBAD
database, the radial velocity is $V_{R} = 34.6$~km s$^{-1}$,
however, according to our measurements, it is systematically
slightly lower which may indicate the possible binarity of the
star. The observed spectrum of HD\,36046 corresponds to a star
with the effective temperature \mbox {$T_\mathrm{eff} = 15000 \pm
250$}~K and the surface gravity $\log g = 4.0 \pm 0.3$.

The paper by \citet{1} provides the references to studies, in
which the attempts have been made to determine the mass of the
star. Two values were obtained: $2.4$ and $3.8~M_{\odot}$. The
second value is close to our data.

According to the paper by~\citet{9}, the star was observed at the
VLT with FORS in order to search for a magnetic field, but the
result was negative. Our observations at the BTA also show no
evidence of a magnetic field stronger than 500 G.

The star is included in the list of Ae/Be Herbig objects in close
OB associations \citep{10}, however, in our spectra in the region
\mbox{4450--4950}~\AA, no evidence of emissions was detected.

\subsubsection{HD\,36313 = V1093\,Orion = BD\,$-00\degr977$ \\= Renson\,9370 = Brown\,014b}

This binary star is variable. The companion is weaker than the
main component by 0.5 mag. and is at the distance of
$0\farcs1$~\citep{11}. \citet{12} found the periodic variability
with the elements $$ \mathrm{HJD(min)} = 2444976.985 + 0.58933\,E.
$$ The fluxes in all filters vary in the phase as a double wave.

\citet {13} first discovered the magnetic field of the star.
According to observations with the Balmer magnetometer, the
variation limits of the longitudinal component of the magnetic
field are from $-1520$ to $1110$~G. Our attempt to measure the
field along the metal lines ended in failure~\citep{14}. More
comprehensive analysis of the data showed that the spectrum of the
star contains two sets of lines differently broadened by rotation.
In the spectrum, there are much more narrow lines belonging to a
cool companion. Apparently, this component of the system does not
have a large-scale field. On the other hand, the H$\beta$ hydrogen
line successfully detects the magnetic field of the main
component. The zero contribution of the companion reduces the
total field by about 30\% which is within the measurement error.
For this reason, we decided to use the original measurements
obtained along the $H\beta$ line to calculate the value $\langle
B_{e}\rangle$ for this star: $$ \langle B_{e} \rangle =
1337~\text{G},~ \sigma = 500,~ \chi^2/n = 7.2. $$ We believe that
the field measurement error $\sigma$ along a single line is 500~G.

\begin{figure*}
\includegraphics[width=12cm]{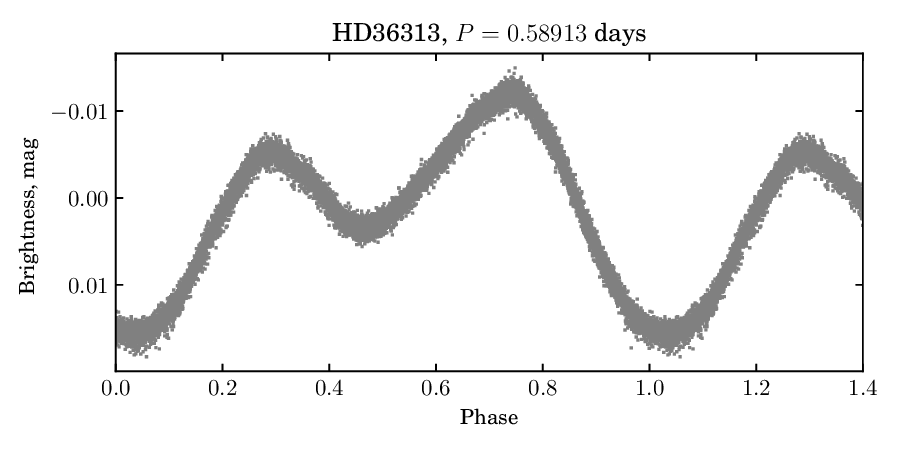}
\caption{Photometric light curve of the magnetic component in the
system HD\,36313 from the TESS observations.} \label{tess_hd36313}
\end{figure*}

The photometric light curve of the star obtained by the TESS
satellite looks like a double wave. On closer examination of the
curve, it appeared that it contains components of at least three
periodic processes. The largest amplitude corresponds to
brightness fluctuations with the period
$P_\mathrm{magn}=0.58913$~days~(Fig.~\ref{tess_hd36313}) which
almost coincides with the data obtained by~\citet{12}. The other
two processes with significantly smaller amplitudes have the
periods $P_{1}=3\fd6729$ and $P_{2}=2\fd987$. Most likely, $P_{1}$
corresponds to the rotation period of the cool component, the
lines of which dominate in the observed spectrum of HD\,36313.

\begin{figure*}
\includegraphics[width=12cm]{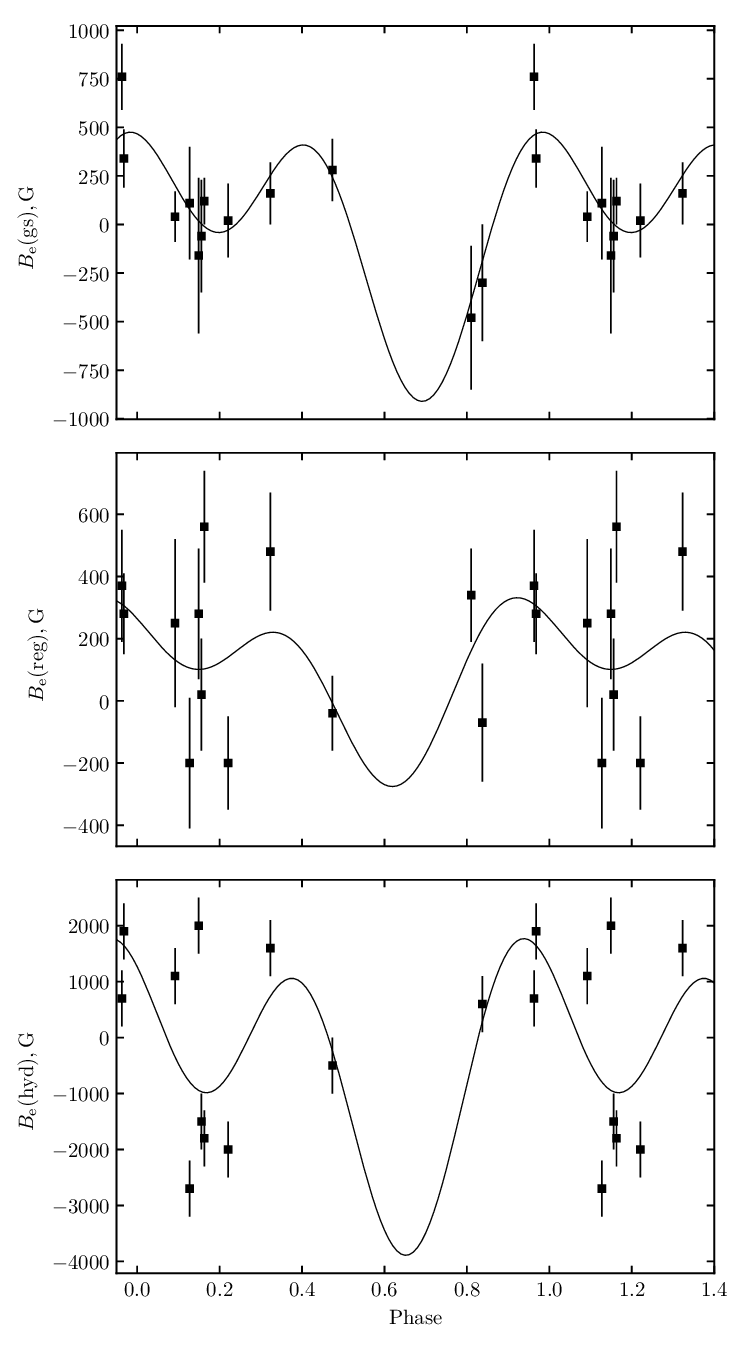}
\caption{Variability curves of the longitudinal magnetic field of
HD 36313 built from our observations.} \label{hd36313_Bzmultipan}
\end{figure*}

Figure~\ref{hd36313_Bzmultipan} gives an idea of the nature of
variations in the longitudinal magnetic field in our measurements.
The period was taken from the value obtained from the TESS
photometry, $T_{0}=2444976.985$. The values $B_{e}$ show large
scattering, but the measurements obtained with the classical
method and along the hydrogen line correlate well with each other.
The phase shift of the photometric and magnetic curves is
noteworthy.

Line broadening in the spectrum corresponds to $v_{e}\sin i_{1} =
160$~km s$^{-1}$ for the major component and to $v_{e}\sin i_{1} =
30$~km s$^{-1}$~for the secondary. We determine the physical
parameters of the major component as follows: $T_{\rm eff} =
13000\pm 480$~K, \mbox {$\log g = 4.0 \pm 0.3$}.

\subsubsection{HD\,36485 = $\delta$\,Orion\,C = BD\,$-00\degr982$ \\= Renson\,9440 = Brown\,020b}

The system $\delta$\,Orion consists of four stars: the primary
component A (HR\,1852 = HD\,36486) is a spectroscopic binary star
itself with \mbox {$m_{V} = 2.23$}~mag., component B is at
$33\farcs0$ from the primary one and has the strength $m_{V} =
14.0$. The star C (HR\,1851 = HD\,36485) with strong helium lines
is at a distance of $51\farcs7$ from A and has the magnitude
$m_{V}$ = 6.85.

HD\,36485~is a well-known magnetic star studied in detail in the
paper by~\citet{15}. The measurements of the magnetic field of the
star are published mainly in the paper by~\citet{16}. The data
show the measurement limits of $B_{e}$ from $-3800$ to $-3100$~G.
\citet{17} found a longitudinal field of $-1850$~G from one
measurement but could not explain the discrepancy with the
Bolender`s results.

We have carried out our own observations of the star. Eleven of
our measurements showed that along the metal lines the field
varies from $-3400$ to $-2300$~G, and from the H$\beta$
line---from $-3100$ to $-1900$~G. The corresponding values of the
root-mean-square longitudinal field obtained with the Babcock
method and from hydrogen are as follows:
\begin{list}{$\bullet$}{
\setlength\leftmargin{10mm} \setlength\topsep{2mm}
\setlength\parsep{0mm} \setlength\itemsep{2mm} }
 \item $\langle B_{e} \rangle = 2724$~G, $\sigma = 261$, $\chi^2/n = 113$ (with the Babcock method),
 \item $\langle B_{e} \rangle = 2156$~G, $\sigma = 132$, $\chi^2/n = 635$ (from hydrogen).
\end{list}

It is known that HD\,36485 is a spectroscopic binary system,
however, speckle interferometric observations reveal one more
component. \citet{18} found a companion at the distance $\rho =
0''\!.327$ fainter than the major star by $\Delta m_\mathrm{V} =
1.5$ mag. . Thus, HD\,36485 is a complex multiple system, the
major component of which has a strong magnetic field of variable
polarity.

Based on our spectra, we have determined the main parameters of
the magnetic star: $v_\mathrm{e}\sin i = 40$ km s$^{-1}$,
$T_\mathrm{eff} = 18000 \pm 250$\,K $\log g = 4.0 \pm 0.3$.

Fairly strong interstellar linear polarization is observed towards
the star: $P = 0.18$\%.

The star was observed with the TESS satellite. However, due to the
complex configuration of the system, it is not possible to
separate the periodic signals in the existing compound light
curve.

\subsubsection{HD 36526 = BD $-01^\circ\,933$ = V 1099 Orion = Renson 9460 = Brown 023b}

The magnetic field of the star with the extrema $B_\mathrm{e}$
from \mbox{$-1000$} to $+3500$\,G was discovered by \citet{13}.
The paper by \citet{14} presents our measurements of its
longitudinal field. \citet{12} found the photometric variability
of the star. According to this paper, the variability is observed
in all filters with the weak secondary minimum and the elements $$
\textrm{HJD(min)} = 2444978.825 + 1.5405\,E. $$

\begin{figure*}
\includegraphics[width=12cm]{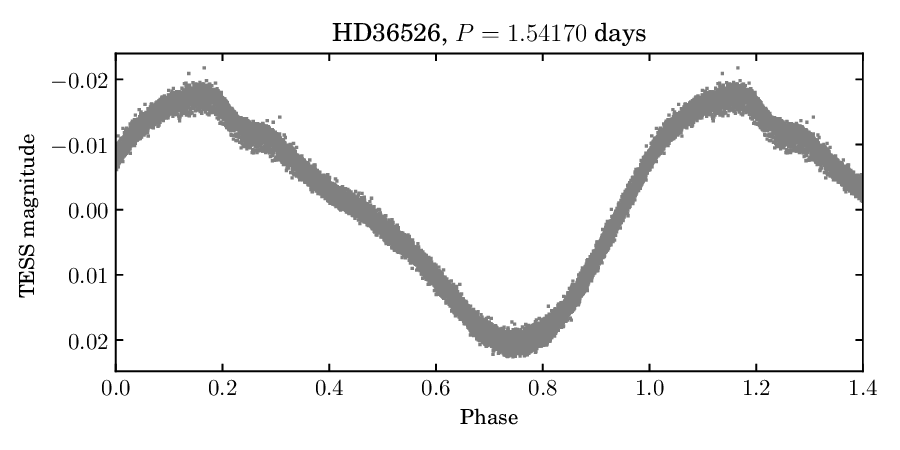}
\caption{Photometric light curve of HD\,36526 from the TESS
observations.} \label{hd36526_tess}
\end{figure*}

\begin{figure*}
\includegraphics[width=12cm]{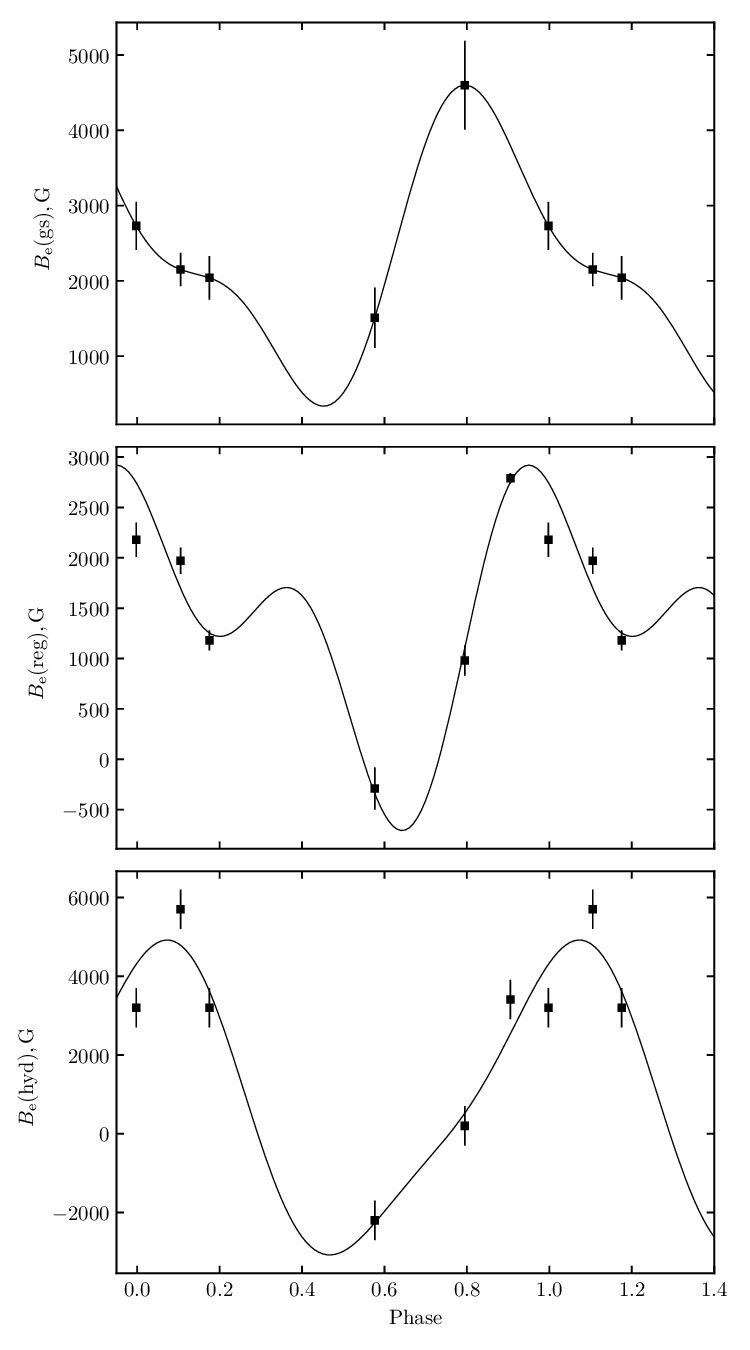}
\caption{Variability curves of the longitudinal magnetic field of
HD\,36526 built from our observations.}
 \label{hd36526_multipan}
\end{figure*}

The light curve obtained with the TESS shows the presence of two
periods: $1\fd54170$ and $1\fd7073$. As is seen, the first value
is close to that obtained by \citet{12} and is the rotation period
of the magnetic star. The light curve of a star phased with a
period of $1\fd54170$ and $T_{0}=2444978.825$ is given in
Fig.~\ref{hd36526_tess}, and Figure~\ref{hd36526_multipan}
presents our measurements of the longitudinal magnetic field with
the same elements.

The character of the spectral variability of HD\,36526 in our data
indicates the possible presence of lines of at least one more
component. Such a component could be a companion detected by
\citet{19} at a distance of $0\farcs15$. The brightness difference
$\Delta m_{V}$ between two stars is only 1.3~mag. Thus, the
spectrum of the companion can have significant impact on our
measurements, as well as on the shape of the light curve of the
magnetic star. According to our measurements, the root-mean-square
field $\langle B_{e} \rangle$ of HD\,36526 is the following:
\begin{list}{$\bullet$}{
 \setlength\leftmargin{10mm} \setlength\topsep{2mm} \setlength\parsep{0mm} \setlength\itemsep{2mm} }
    \item $\langle B_{e} \rangle = 2801$~G, $\sigma = 384$, $\chi^2/n = 56.8$ (with the Babcock method),
    \item $\langle B_{e} \rangle = 1695$~G, $\sigma = 137$, $\chi^2/n = 539.3$ (with the regression method).
\end{list}

Preliminary analysis of the spectra gives the following
fundamental parameters of the magnetic star: $T_\mathrm{eff}=16000
\pm 210$~K, \mbox {$\log g = 4.0 \pm 0.3$}. The rotational line
broadening corresponds to $v_{e}\sin i = 50$~km s$^{-1}$. There
are no measurements of radial velocities in the literature.
Measurements of the spectra obtained with the BTA give the value:
$V_{R} = 30$~km s$^{-1}$.

\subsubsection{HD\,36668 = BD\,$+00\degr1113$ = V1107\,Orion \\= Renson\,9560 = Brown\,031b}

For the first time, the magnetic field of the star was measured by
\citet{13} from observations with a Balmer magnetometer. The
obtained $B_{e}$ values ranged from $-2100$ to $+2000$~G. We
carried out eight observations of the star. Measurements of the
Zeeman effect in the metal and hydrogen lines confirm the presence
of a strong field, but its variation limits are much larger than
those published by \citet{4}.

Given the available data, the root-mean-square field values are as
follows:
\begin{list}{$\bullet$}{
\setlength\leftmargin{10mm} \setlength\topsep{2mm}
\setlength\parsep{0mm} \setlength\itemsep{2mm}}
   \item $\langle B_{e} \rangle = 1892$~G, $\sigma = 451$, $\chi^2/n = 37.6$ (with the Babcock method),
   \item $\langle B_{e} \rangle = 953$~G, $\sigma = 105$, $\chi^2/n = 203.5$ (with the regression method).
\end{list}

This star is a photometric variable. \citet{12} gives the
following light curve elements: $$ \textrm{HJD(min)} = 2444988.496
+ 2.1211\,E. $$

\begin{figure*}
\includegraphics[width=12cm]{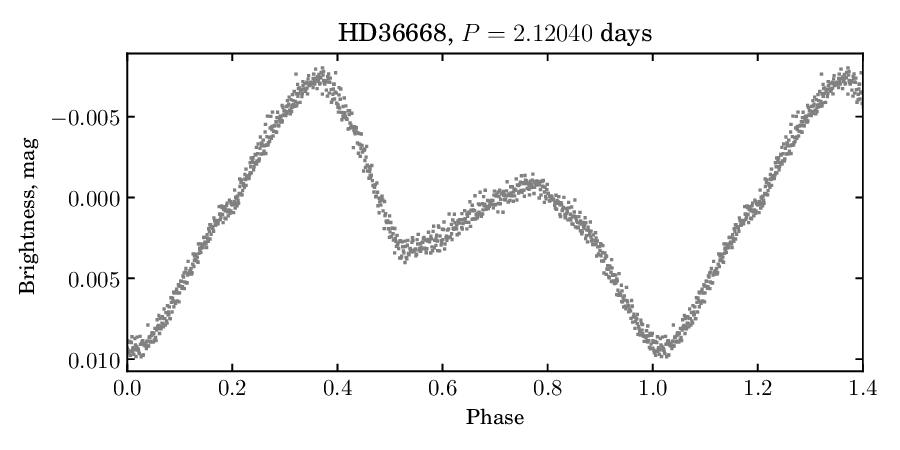}
\caption{Photometric light curve of HD\,36668 from the  TESS
observations.} \label{hd36668_tess}
\end{figure*}

The brightness variation occurs in the shape of a double wave, in
which the secondary maximum is almost as deep as the main one. The
star was observed with the TESS satellite, but its photometry was
conducted separately in the CDIPS
survey~\citep{2019ApJS..245...13B}. Analysis of the data cleared
of instrumental trends gives the variability period \mbox
{$P=2\fd1204$} close to that given above from the paper
by~\citet{12}. The final light curve of the star has a complex
shape and is shown in Fig.~\ref{hd36668_tess}.

We estimated the fundamental parameters: $v_{e}\sin i =60$~km
s$^{-1}$, $T_\mathrm{eff} = 13500 \pm 250$~K, $\log g = 4.0 \pm
0.4$. Based on these data, as well as the position of the star in
space, the luminosity and mass of the star can be found. These
values for HD\,36668 are equal to $\log \dfrac{L}{L_{\odot}} =
2.4$ and $\dfrac{M}{M_{\odot}} = 3.7$, respectively. The paper by
\citet{1} gives two mass estimates for HD\,36668 equal to
$3.8~M_\odot$, which coincides with the given results.

The paper by \citet{20} claims that HD\,36668 does not belong to
the association but to the stellar flow in the Orion and is closer
to the observer. However, the parallax obtained during the GAIA
mission disproves this claim. The value $\pi = 2.36$~mas
corresponds to the distance $d = 424$~pc, and this is the distance
to the center of subgroup~1b of the association \citep{1}.

\citet{10} included HD\,36668 into the catalog of Ae/Be Herbig
stars in the nearby OB associations, but in our spectra in the
region of \mbox{4450--4950} no evidence of Ae/Be stars was found.

\subsubsection{HD\,36955 = BD\,$-01\degr955$ = Renson\,9740 \\= Brown\,052b}

We discovered the magnetic field of this star with the
BTA~\citep{21}. The SIMBAD database indicates that HD\,36955~is a
binary or multiple system. The paper by~\citet{22} says about the
presence of a companion with the brightness $m_{V} = 11$~mag. at a
distance of $1\farcs5$.

\begin{figure*}
\centering\includegraphics[width=12cm]{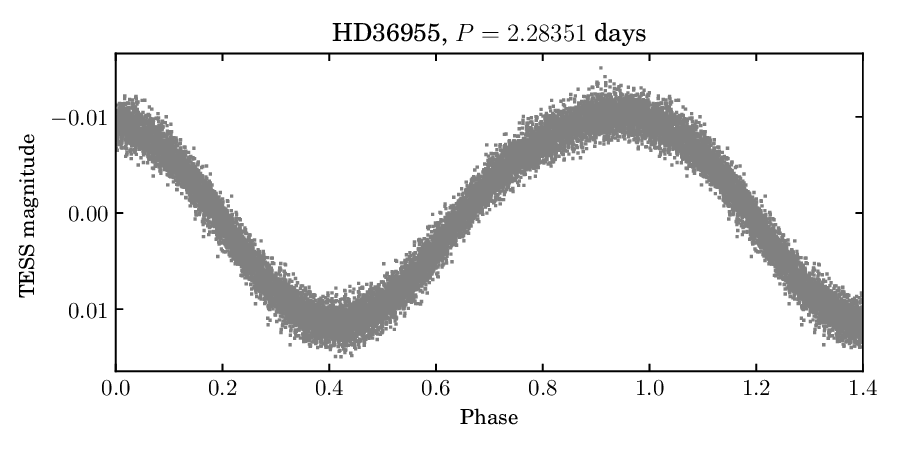}
\caption{Photometric light curve of HD\,36955 from the TESS
observations.} \label{hd36955_tess}
 \end{figure*}

\begin{figure*}
\includegraphics[width=12cm]{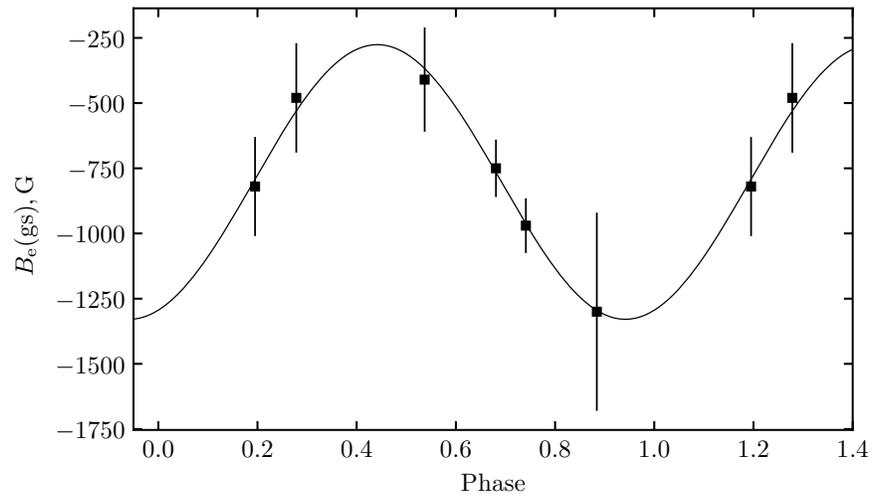}
\caption{Variability curve of the longitudinal magnetic field of
HD\,36955 obtained from the measurements of the metal lines.}
\label{hd36955_multipan}
\end{figure*}

\citet{23} found the rotation period of the star: $P =
2\fd284965$. The light curve, built from the photometry performed
with the TESS satellite, is of a simple
shape~(Fig.~\ref{hd36955_tess}). The refined value of the
photometric variability period is equal to $2\fd283506$. Both
these values are in poor agreement with the values of the
longitudinal magnetic field. Analysis of our measurements of
$B_{e}$ testifies in favor of a slightly longer period:
$2\fd875108$~(Fig.~\ref{hd36955_multipan}). The reasons for these
discrepancies remain to be understood.

The root-mean-square longitudinal magnetic field of the star
measured with the Babcock and regression methods is as follows:
 \begin{list}{$\bullet$}{
 \setlength\leftmargin{10mm} \setlength\topsep{2mm} \setlength\parsep{0mm} \setlength\itemsep{2mm}}
   \item $\langle B_{e} \rangle = 842$~G, $\sigma = 219$, $\chi^2/n = 28.2$,
   \item $\langle B_{e} \rangle = 708$~G, $\sigma = 90$, $\chi^2/n = 93.9$.
 \end{list}

The fundamental parameters of HD\,36995 that we determined are:
$v_{e}\sin i = 26$~km s$^{-1}$, $T_\mathrm{eff} = 10800\pm250$~K,
$\log g = 4.2\pm0.5$.

\subsubsection{HD\,37140 = V1130\,Orion = BD\,$-00\degr1018$ \\= Renson\,9910 = Brown\,063b}

The magnetic field of the star was found by~\citet{13}. In his
measurements, the longitudinal field component ranged from $-1050$
to $+400$~G. \citet{12} found the photometric variability of the
star with the elements $$ \mathrm{HJD(min)} = 2444978.036 +
2.7088\,E. $$ The light curves in all filters are sinusoidal.
However, there is another probable rotation period in the
literature: $0\fd611465$~\citep{23}.

\begin{figure*}
\includegraphics[width=12cm]{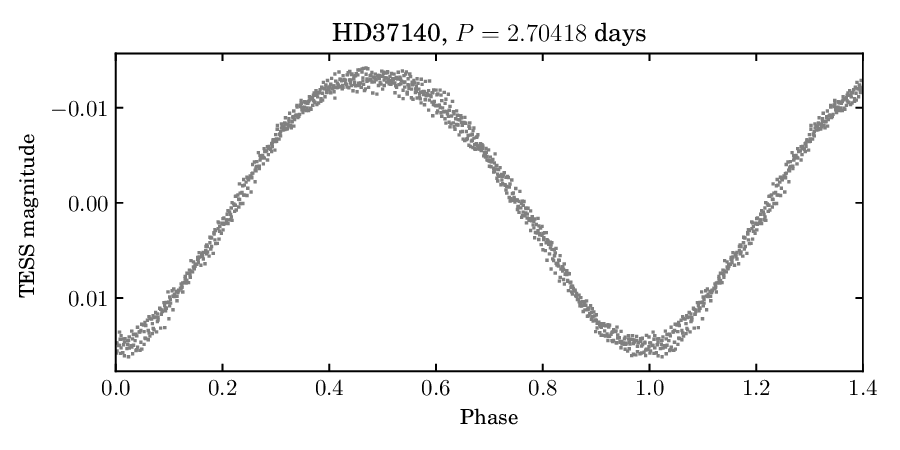}
\caption{Photometric light curve of HD\,37140 from the TESS
observations.} \label{hd37140_tess}
 \end{figure*}

The photometry of the star based on the TESS images was carried
out in the CDIPS project. The light curve cleared of trends is
perfectly phased with a period of $P=2\fd704179$ which is close to
North`s value. \citet{22} argues that HD\,37140~is a triple system
with components of the A7 and F--K spectral types at the distances
$d = 23$~au and 50~au. The hotter companion can be a pulsating
star of the $\delta$\,Sct type, as there are the signs of specific
pulsations of a small amplitude in the TESS photometry (see
Fig.~\ref{hd37140_tess}).

\begin{figure*}
\includegraphics[width=12cm]{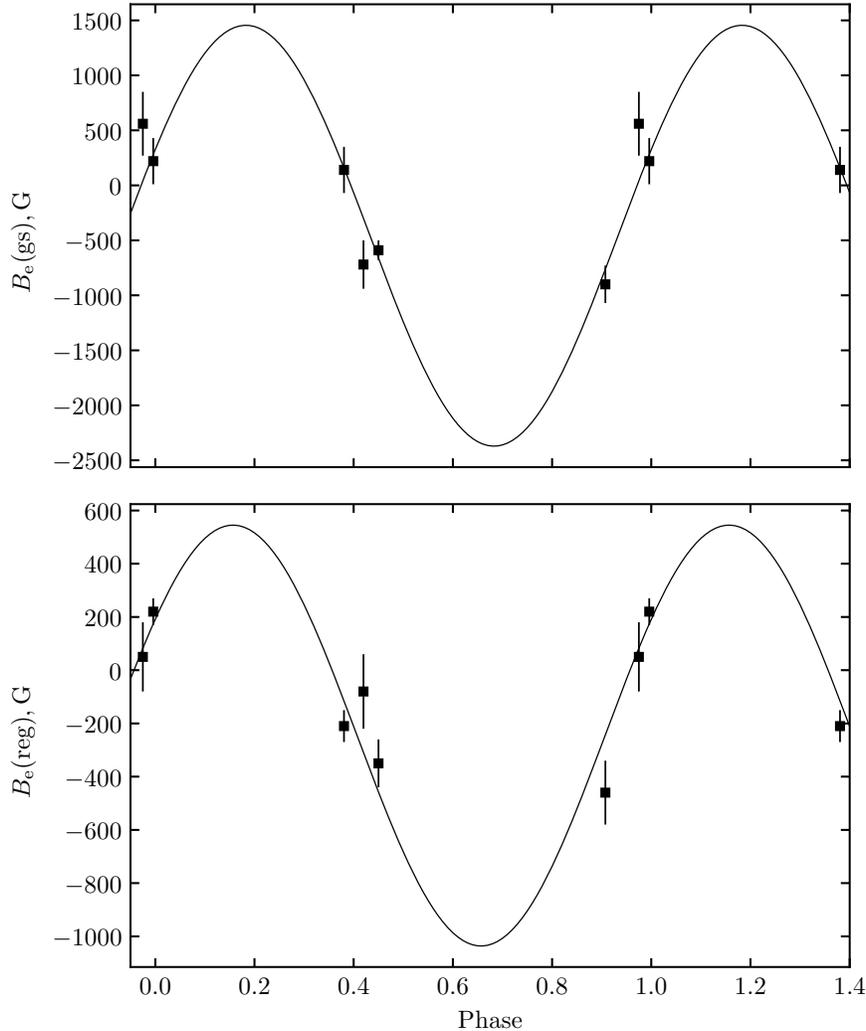}
\caption{Variability curves of the longitudinal magnetic field of
HD\,37140 obtained from the measurements of metal lines with the
Babcock method and the regression method.}
\label{hd37140_multipan}
\end{figure*}

The star HD\,37140 is magnetic: our longitudinal field
measurements with two methods are given in
Fig.~\ref{hd37140_multipan}. The root-mean-square longitudinal
magnetic fields from our data depending on the method used~(the
modified Babcock method, the regression method) are the following:
\begin{list}{$\bullet$}{
\setlength\leftmargin{10mm} \setlength\topsep{2mm}
\setlength\parsep{0mm} \setlength\itemsep{2mm}}
  \item $\langle B_{e} \rangle = 585$~G, $\sigma = 207$, $\chi^2/n = 14.0$,
  \item $\langle B_{e} \rangle = 270$~G, $\sigma = 107$, $\chi^2/n = 9.6$.
 \end{list}

Having analyzed the spectra, we found the following parameters of
the star: $v_{e}\sin i = 30$~km s$^{-1}$, \mbox {$T_\mathrm{eff} =
13500\pm240$}~K, $\log g = 3.7\pm0.4$.

\subsubsection{HD\,37149 = HIP\,26319 = Renson\,9920 \\= Brown\,065b}

In the catalog of chemically peculiar stars \citep{6}, HD\,37149
is named as \mbox {He-wk}, but the classification can be wrong:
there is information in the literature that HD\,37149~is a Be
star. Despite the fact that there is no evidence of emission in
our spectra obtained in the region of 4450--4950~\AA, we will
consider it a non-magnetic \mbox {Be star}. This assumption is
also supported by unsuccessful attempts to detect the magnetic
field of the star (with the Babcock method and by regression,
respectively):
 \begin{list}{$\bullet$}{
 \setlength\leftmargin{10mm} \setlength\topsep{2mm} \setlength\parsep{0mm} \setlength\itemsep{2mm}}
   \item $\langle B_{e} \rangle = 1826$~G, $\sigma = 2238$, $\chi^2/n = 0.4$,
   \item $\langle B_{e} \rangle = 205$~G, $\sigma = 170$, $\chi^2/n = 2.1$.
 \end{list}

\begin{figure*}
\includegraphics[width=17.8cm]{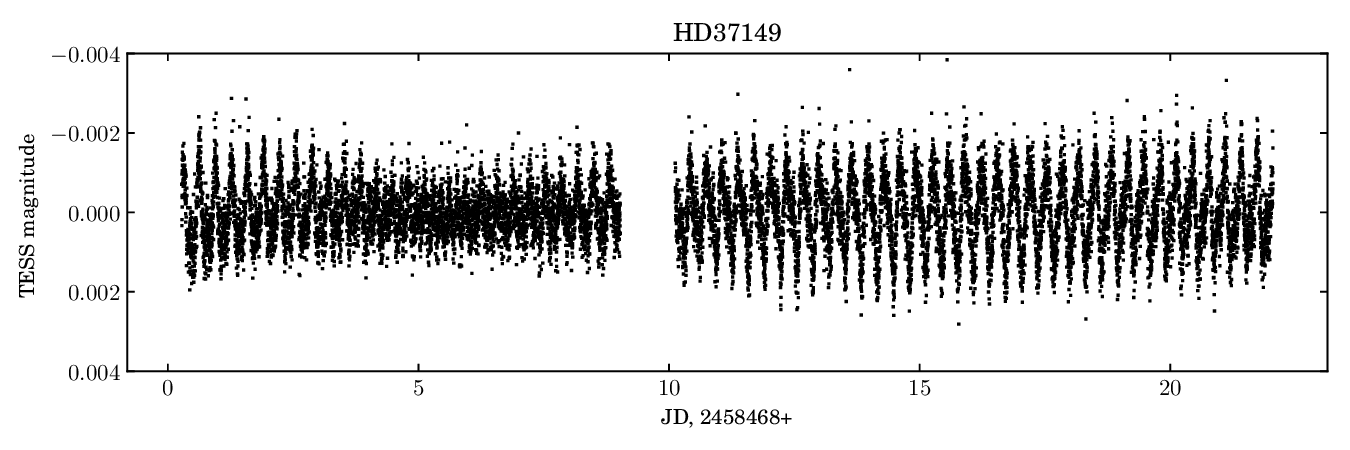}
\caption{Photometric light curve of HD\,37149 from the TESS
observations.} \label{hd37149_tess}
\end{figure*}

The stellar photometry conducted at the TESS satellite shows the
multiperiodic small-amplitude
variability~(Fig.~\ref{hd37149_tess}). Possible oscillation
periods: ($P_{1}=0\fd3196$ and \mbox {$P_{2}=0\fd3245$}) are also
typical for \mbox {Be stars} \citep{2009CoAst.158..194N}. Since,
HD\,37149 is not a chemically peculiar star, we exclude it from
our further consideration.

\subsubsection{HD\,37235 = BD\,$-00\degr1023$ = Renson\,9960 \\= Brown\,069b}

This star has not been previously studied as for magnetic field
searches. The available spectra show three more or less strong
lines broadened by rotation. For this reason, the accuracy of the
longitudinal magnetic field measurement is very low. The results
of measurements of six Zeeman spectra indicate that there is no
longitudinal magnetic field stronger than 1 kG. No evidence of the
Zeeman signature is observed in either the metal or in the
hydrogen lines. The root-mean-square magnetic field depending on
the method is as follows:
\begin{list}{$\bullet$}{
\setlength\leftmargin{10mm} \setlength\topsep{2mm}
\setlength\parsep{0mm} \setlength\itemsep{2mm}}
  \item $\langle B_{e} \rangle = 4204$~G, $\sigma = 2889$, $\chi^2/n = 1.2$ (the modified Babcock method),
  \item $\langle B_{e} \rangle = 227$~G, $\sigma = 170$, $\chi^2/n = 2.3$ (the regression method).
  \end{list}

\begin{figure*}
\includegraphics[width=12cm]{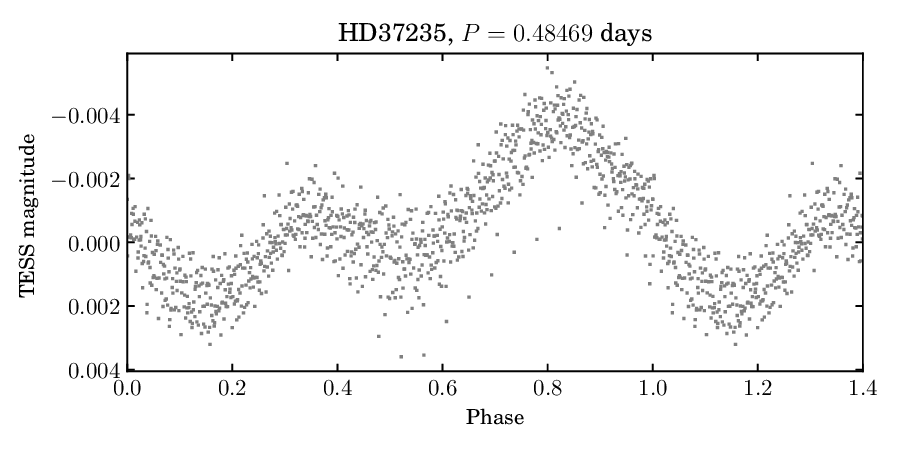}
\caption{Photometric light curve of HD\,37235 from the TESS
observations.} \label{hd37235_tess}
\end{figure*}

The negative result of magnetic field searches for HD\,37235
should not be considered a sign that the star is not chemically
peculiar. The photometric light curve of HD\,37235 is available in
the CDIPS survey based on the analysis of images obtained with the
TESS satellite. The periodogram clearly shows the signal
corresponding to the period of the star variability
\mbox{$P=0\fd48469$}. According to this parameter, HD\,37235~is
one of the fastest rotators with the photometric variability. The
brightness of the star varies within narrow limits in the shape of
a double wave typical of \mbox {CP
stars}~(Fig.~\ref{hd37235_tess}).

We determined the physical parameters of the star: $v_{e}\sin i =
320$~km s$^{-1}$, \mbox {$T_\mathrm{eff} = 13500\pm300$}~K, $\log
g = 4.0\pm0.3$.

\subsubsection{HD\,37321 = HIP\,26439 = Renson\,10000 \\= Brown\,075b}

This massive star~$(5.2~M_{\odot}$) is the major component of the
binary system ADS\,4222AB with a companion at a distance of
$0\farcs8$~\citep{1}. In the wavelength range of the spectra that
we obtained, few lines are observed. The fast rotation of the star
results not only in broadening of its lines but also significantly
reduces the accuracy of the magnetic field measurement.

None of the five measurements of the longitudinal field showed any
significant result: we could not find a magnetic field. The
root-mean-square $\langle B_{e} \rangle$ values, that we found
with the modified Babcock method and the regression method, are
the following:
\begin{list}{$\bullet$}{
\setlength\leftmargin{10mm} \setlength\topsep{2mm}
\setlength\parsep{0mm} \setlength\itemsep{2mm}}
  \item $\langle B_{e} \rangle = 426$~G, $\sigma = 439$, $\chi^2/n = 0.9$,
  \item $\langle B_{e} \rangle = 290$~G, $\sigma = 202$, $\chi^2/n = 1.4$.
  \end{list}

Our estimates of the physical parameters of the star are:
$v_{e}\sin i=130$~km s$^{-1}$, \mbox {$T_\mathrm{eff}=15000 \pm
350$}~K, \mbox {$\log g = 4.1\pm 0.4$}, $\log
\dfrac{L}{L_{\odot}}\!=\!3.3$, $\dfrac{M}{M_{\odot}}\!=\!5.8$,
$\dfrac{R}{R_{\odot}}\!=\!4.6$.

The mass is in good agreement with the estimate from the paper
by~\citet{1}. In TESS photometry (the CDIPS survey), the star
exhibits weak multiperiodic variability typical of massive
pulsating stars.

\subsubsection{HD\,37333 = BD\,$-02\degr1319$ = Renson\,10010 \\= Brown\,077b}

HD\,37333~is a new star, the member of the $\sigma$\,Orion
cluster. The silicon lines in the star spectrum are strong.

\citet{9} observed HD\,37333 with FORS1 VLT, although, no magnetic
field was found. We found a magnetic field for the first time, but
the results have not been previously published. The
root-mean-square longitudinal field $\langle B_{e} \rangle$ is as
follows from our six observations:
\begin{list}{$\bullet$}{
\setlength\leftmargin{10mm} \setlength\topsep{2mm}
\setlength\parsep{0mm} \setlength\itemsep{2mm}}
  \item $\langle B_{e} \rangle = 869$~G, $\sigma = 246$, $\chi^2/n = 19.4$,
  \item $\langle B_{e} \rangle = 433$~G, $\sigma = 120$, $\chi^2/n = 14.6$.
\end{list}

As usually, the root-mean-square field measured with the
regression method is smaller than that measured with the classical
method.

\begin{figure*}
\includegraphics[width=12cm]{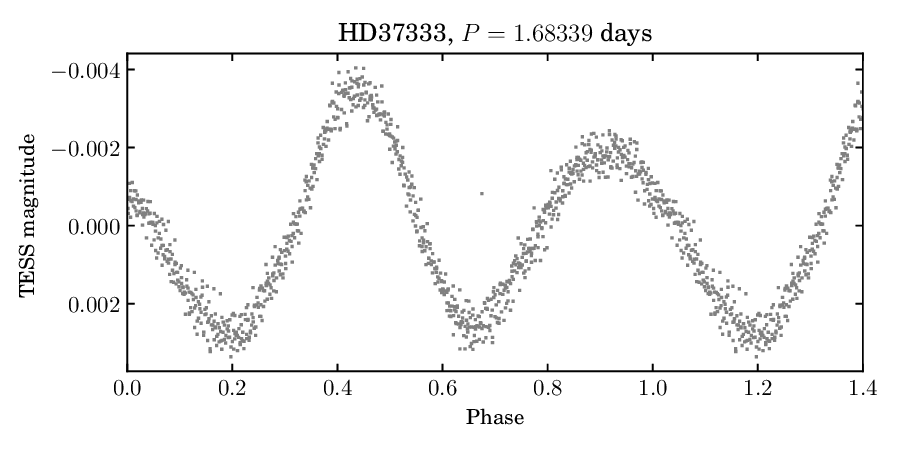}
\caption{Photometric light curve of HD\,37333 from the TESS
observations.} \label{hd37333_tess}
\end{figure*}

In the catalog by \citet{24}, the rotation period of the star $P =
5\fd612112$ is given; however, our measurements of $B_{e}$ do not
agree with it. Analysis of the light curve obtained in the CDIPS
survey based on the TESS observations shows that the current
rotation period is $1\fd68339$. With this period, the variations
in the star brightness occurs in the form of a double wave with
two minima of the same depth~(Fig.~\ref{hd37333_tess}). Magnetic
measurements phased with a specified photometric period fall into
the phase range of 0.45--0.85. These data are insufficient for any
conclusions about the nature of the magnetic variability.

Based on the data available, we found the following fundamental
parameters of the star: $v_{e}\sin i = 50$~km s$^{-1}$,
$T_\mathrm{eff} = 12000\pm370$~K, \mbox {$\log g = 4.5\pm0.3$}.

\subsubsection{HD\,37479 = $\sigma$\,Orion\,E = BD\,$-02\degr1327$ \\= Renson\,10080 = Brown\,086b}

HD\,37479~is a well-studied magnetic peculiar star with strong
helium lines. \citet{16} performed 22 measurements of the star
longitudinal magnetic field with the Landstreet`s Balmer
magnetometer. The authors obtained an approximately sinusoidal
curve. The purpose of our measurements is to obtain data for all
magnetic stars in a uniform manner in a homogeneous system in
order to be able to compare the results obtained from the H$\beta$
hydrogen line and from the metal lines. Other details of our
research are given in the paper by \citet{1}.

We confirm that star has a very strong magnetic field. The
root-mean-square longitudinal fields obtained with the Babcock
method and with the regression method are as follows:
\begin{list}{$\bullet$}{
\setlength\leftmargin{10mm} \setlength\topsep{2mm}
\setlength\parsep{0mm} \setlength\itemsep{2mm}}
  \item $\langle B_{e} \rangle = 3040$~G, $\sigma = 853$, $\chi^2/n = 29.2$,
  \item $\langle B_{e} \rangle = 1307$~G, $\sigma = 324$, $\chi^2/n = 46.6$.
\end{list}

There is a very large difference in the results obtained with
these two methods.

The fundamental parameters of the star are the following:
$v_{e}\sin i = 150$~km s$^{-1}$, $T_\mathrm{eff} = 21000\pm550$~K,
$\log g = 3.5\pm0.4$.

\subsubsection{HD\,37525 = BD\,$-02\degr1328$ = Renson\,10110 \\= Brown\,088b}

This star is presented in the SIMBAD database as a young object.
The binary system HD\,37525AB belongs to the $\sigma$\,Orion
cluster. The catalog by \citet{6} classifies the star as peculiar
with weak helium abundance; however, in the spectrum, the helium
4471 \AA\  line is significantly stronger than the Mg\,II 4481~\AA
line. This means that the helium abundance is not that small. One
cannot exclude that HD\,37525~is a normal star of a corresponding
spectral type.

There are no data on any measurements of its magnetic field in the
literature. In five observations at the BTA with the Zeeman
analyzer, no longitudinal field with an upper limit of 500 G was
found also. The root-mean-square field $\langle B_{e} \rangle$ is
as follows from our observations:
\begin{list}{$\bullet$}{
\setlength\leftmargin{10mm} \setlength\topsep{2mm}
\setlength\parsep{0mm} \setlength\itemsep{2mm}}
  \item $\langle B_{e} \rangle = 1362$~G, $\sigma = 1845$, $\chi^2/n = 0.5$ (with the Babcock method),
  \item $\langle B_{e} \rangle = 127$~G, $\sigma = 192$, $\chi^2/n = 0.3$ (with the regression method).
\end{list}

Fundamental parameters HD\,37525 that we obtained are as follows:
$v_{e}\sin i = 160$~km s$^{-1}$, $T_\mathrm{eff} = 17000\pm270$~K,
$\log g = 4.1\pm0.3$.

\subsubsection{HD\,37633 = BD\,$-02\degr1332$ = Renson\,10130 \\= Brown\,093b}

We found the magnetic field of this star in 2013, but the
measurements have not been previously published. \citet{9}
obtained one measurement of the longitudinal field with the FORS1
VLT: \mbox {$B_{z} = 440\pm200$}~G.

\citet{12} found periodic photometric variability with the
elements $$ \mathrm{HJD(min)} = 2444983.923 + 1.5718\,E. $$

\begin{figure*}
\includegraphics[width=12cm]{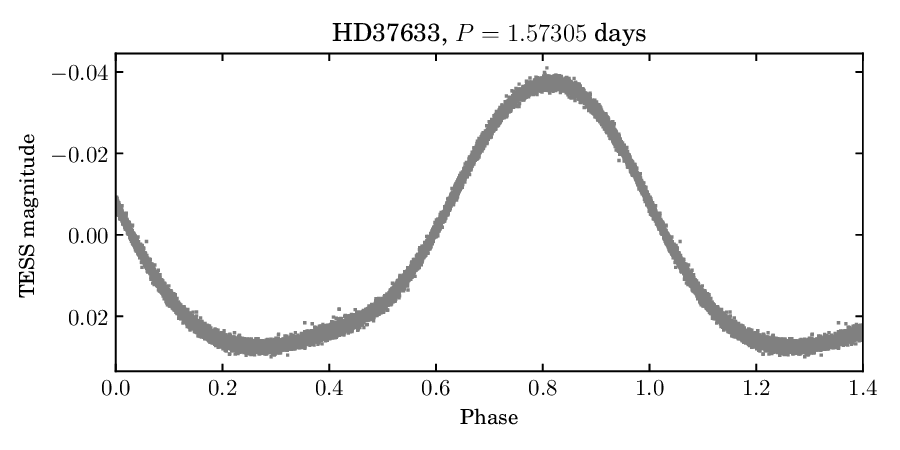}
\caption{Photometric light curve of HD\,37633 from the TESS
observations.} \label{hd37633_tess}
\end{figure*}

\begin{figure*}
\includegraphics[width=12cm]{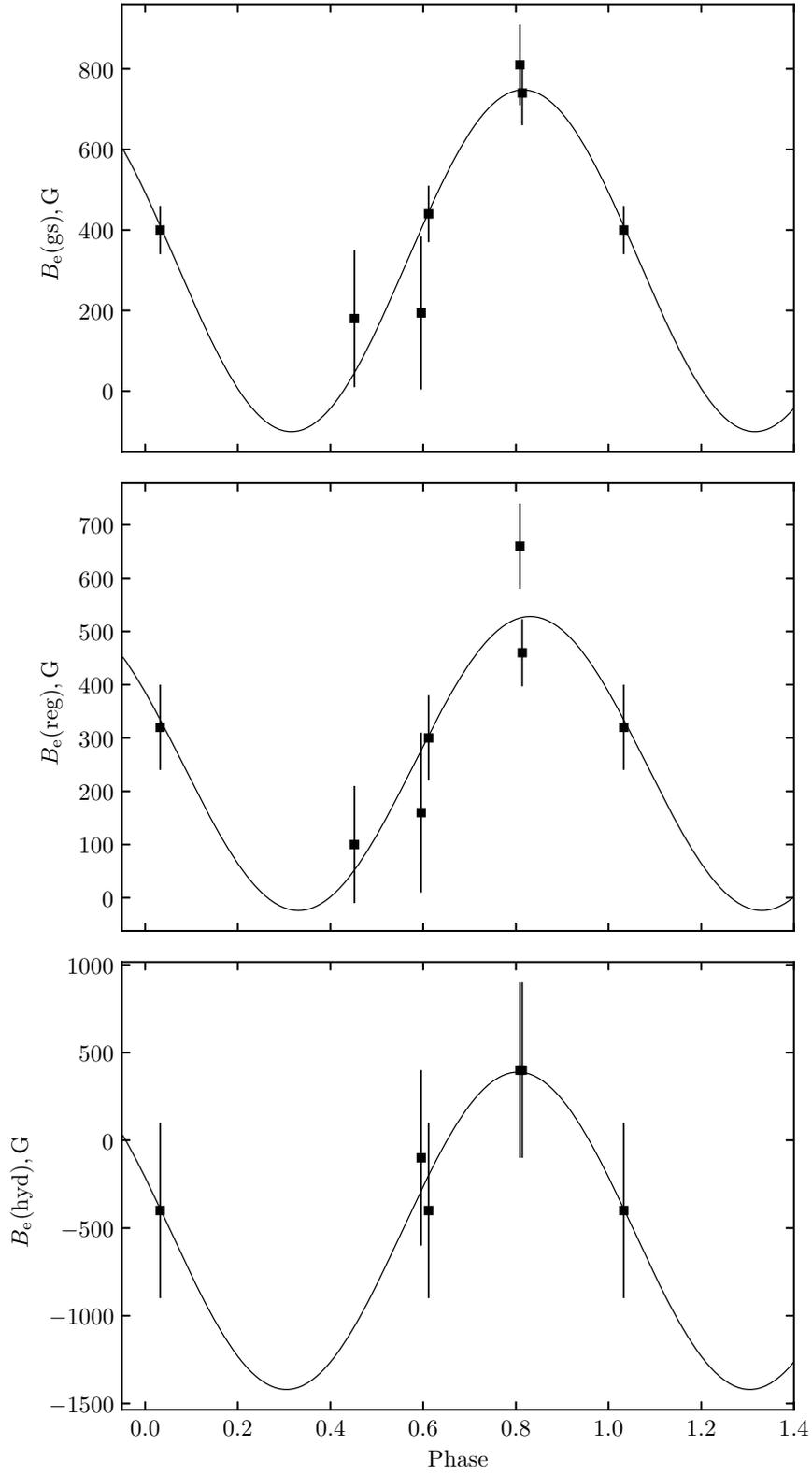}
\caption{Variability curves of the longitudinal magnetic field of
HD\,37633 obtained from the measurements of the metal lines and
hydrogen.} \label{hd37633_multipan}
\end{figure*}

The light curve of a star, as observed by TESS, has two harmonics
with the flatter minimum.(Fig.~\ref{hd37633_tess}). The
variability period practically coincides with the period from the
paper by \citet{12}: \mbox {$P = 1\fd57305$}. Our longitudinal
magnetic field measurements are in good agreement with this
value~(Fig.~\ref{hd37633_multipan}).

The root-mean-square longitudinal magnetic field indicates that
star is magnetic:
\begin{list}{$\bullet$}{
\setlength\leftmargin{10mm} \setlength\topsep{2mm}
\setlength\parsep{0mm} \setlength\itemsep{2mm}}
  \item $\langle B_{e} \rangle = 520$~G, $\sigma = 121$, $\chi^2/n = 41.2$ (with the Babcock method),
   \item $\langle B_{e} \rangle = 382$~G, $\sigma = 97$, $\chi^2/n = 25.5$ (with the regression method).
\end{list}

According to various sources, the star is included in the
$\sigma$\,Orion and Collinder\,70 clusters.

We found the following parameters of the star: $v_{e}\sin i =
35$~km s$^{-1}$, $T_\mathrm{eff} = 13000\pm250$~K, \mbox
{$\log g = 4.0\pm0.4$}.

\subsubsection{HD\,37776 = V901\,Orion = BD\,$-01\degr1005$ \\= Renson\,10190 = Brown\,104b}

HD\,37776~is a well-known magnetic chemically peculiar star
studied many times by various authors, including the authors of
this paper, over decades. The star has an extremely strong field
of a complex non-dipole configuration~\citep{25}. Despite the
unconditional progress in the study of this unique object, in our
opinion, a satisfactory magnetic model of HD\,37776 has not yet
been built. New high-accuracy photometric observations with the
TESS mission raised new issues rather than brought researchers
closer to unravelling the HD\,37776 phenomenon.

During the winter season of 2019/2020, we carried out fourteen
observations of the star with the Zeeman analyzer. The lines in
the spectrum of HD\,37776 are of a very complex shape; there is an
extremely strong circular polarization in the lines caused by the
Zeeman effect. In this case, the lines of different chemical
elements behave differently which results in strong scattering of
field measurements. The use of the regression method appeared to
be ineffective due to a strong field of complex geometry.

The root-mean-square $\langle B_{e} \rangle$ values found from the
measurements of metal lines and hydrogen are the following:
 \begin{list}{$\bullet$}{
 \setlength\leftmargin{10mm} \setlength\topsep{2mm} \setlength\parsep{0mm} \setlength\itemsep{2mm}}
   \item $\langle B_{e} \rangle = 7285$~G, $\sigma = 2686$, $\chi^2/n = 17.2$ (with the Babcock method),
   \item $\langle B_{e} \rangle = 8644$~G, $\sigma = 500$, $\chi^2/n = 298.9$ (by hydrogen).
 \end{list}

Table~\ref{tab3} presents the results of measurements of the
longitudinal field for four elements: H$\beta$, Mg\,II (4481~\AA),
He\,I (4471, 4713, 4922~\AA), and Si\,III (4552, 4567, 4574~\AA).
For helium and silicon, the average values were taken along the
indicated three lines.

\begin{table*}[]
\caption{Longitudinal field $B_{e}$ measurements of the star
HD\,37776 by specified elements} \label{tab3}
\medskip
\begin{tabular}{c|c|c|r@{$\,\pm\,$}l|r@{$\,\pm\,$}l}
\hline
JD, (2450000+) & $B_{e}$(H$\beta)$, kG & $B_{e}$(Mg\,II), kG & \multicolumn{2}{c|}{$B_{e}$(He\,I), kG} & \multicolumn{2}{c}{$B_{e}$(Si\,III), kG} \\
\hline
8777.441  & $6.6$   & $-12.2$ &  $11.1$ & $4.0$  &  $30.5$  & $3.0$  \\
8777.553  & $-3.1$  & $-9.0$  &  $-6.8$ & $1.9$  &  $-5.3$  & $2.2$  \\
8778.460  & $7.0$   & --      &  $5.2$  & $5.1$  &  \multicolumn{2}{c}{--} \\
8778.578  & $1.5$   & $-10.9$ &  $0.8$  & $1.1$  &  $-20.8$ & $3.0$  \\
8799.396  & $6.9$   & $-5.0$  &  $9.0$  & $1.2$  &  $-15.5$ & $1.4$  \\
8801.555  & $7.8$   & $-9.5$  &  $7.1$  & $9.3$  &  $-18.5$ & $1.6$  \\
8805.369  & $-5.3$  & $-4.8$  &  $-5.3$ & $0.8$  &  $-7.2$  & $0.1$  \\
8805.559  & $12.3$  & --      &  $16.0$ & $11.5$ &  $-5.5$  & $2.2$  \\
8830.315  & $10.8$  & $-7.4$  &  $4.7$  & $7.3$  &  $-13.9$ & $2.3$  \\
8830.478  & $-4.8$  & $-6.8$  &  $-5.8$ & $1.1$  &  $-7.6$  & $1.2$  \\
8834.418  & $4.3$   & $-11.5$ &  $4.5$  & $2.2$  &  $27.6$  & $0.9$  \\
8834.511  & $-2.5$  & $-4.6$  &  $-4.6$ & $0.4$  &  $-5.2$  & $3.5$  \\
8855.184  & $-0.7$  & $-4.9$  &  $-2.9$ & $0.5$  &  $-7.1$  & $1.0$  \\
8857.255  & $-15.6$ & $12.7$  &  $-6.7$ & $7.7$  &  $20.0$  & $2.8$  \\
\hline
\end{tabular}
\end{table*}

\begin{figure*}
\includegraphics[width=12cm]{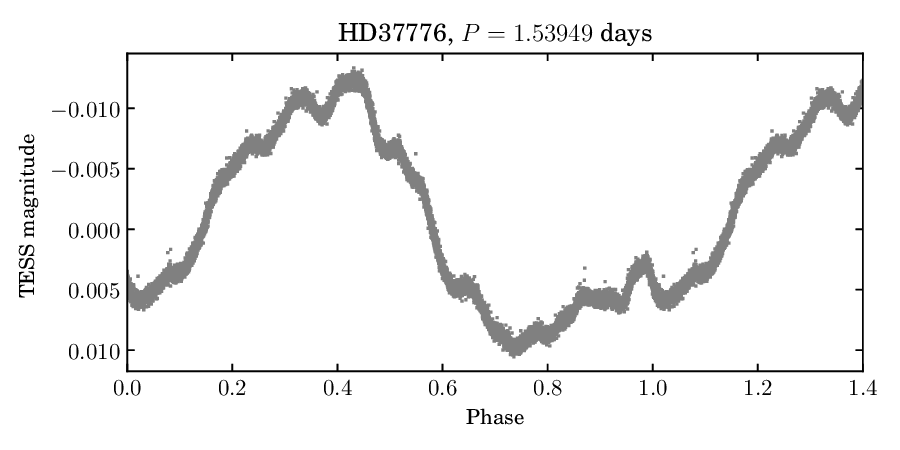}
\caption{Photometric light curve of HD\,37776 from the TESS
observations.} \label{hd37776_tess}
\end{figure*}

\begin{figure*}
\includegraphics[width=12cm]{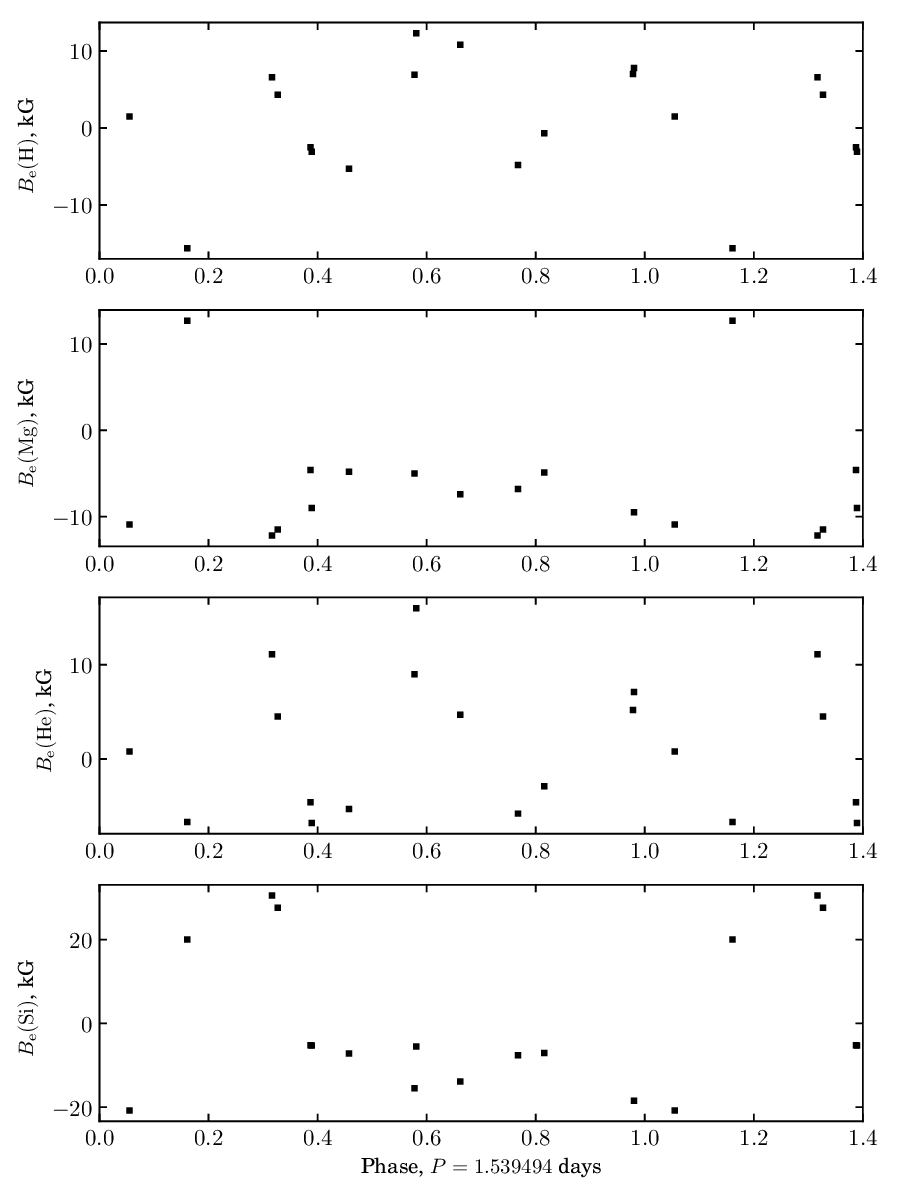}
\caption{Variations of the longitudinal magnetic field of
HD\,37776 with phase measured for different chemical elements.}
\label{hd37776_mfield}
\end{figure*}

The longitudinal magnetic field exhibits very different behavior
depending on the element. As the rotation period, we took the
value $1\fd539494$ determined from the TESS
photometry~(Fig.~\ref{hd37776_tess}). Figure~\ref{hd37776_mfield}
shows variations of the longitudinal magnetic field depending on
the rotation period phase.

The line profiles in the available spectra often bifurcate
indicating Zeeman splitting in the field within the order of 70
kG. It can be clearly seen that the brightness extrema coincide
with the extrema of the magnetic field. In this case, the field
varies in different ways for different elements. For example, the
field along the helium and silicon lines changes in anti-phase. In
this paper, we publish only the first results. More detailed
analysis of the HD\,37776 field is to be performed, but the
presence of a very large (many dozen kG) and complex, unparalleled
field is obvious.

The star has the following physical parameters: $v_{e}\sin i =
80$~km s$^{-1}$, $T_\mathrm{eff} = 22000\pm350$~K, $\log g =
3.7\pm0.6$. The effective temperature found that we found is in
good agreement with plenty of published data. The mass of the
star, according to different sources, ranges from $6.4$ to
$10~M_{\odot}$~\citep{1}.

\subsubsection{HD\,290665 = BD\,$-00\degr1008$ = Renson\,9760 \\= Brown\,128b}

We detected the magnetic field of this star at the 6-m BTA
telescope. The longitudinal component field varies from $-3700$ to
$3900$~G. One more measurement field of this star was performed by
\citet{30} with the VLT: $B_{l} = -1664\pm44$~G.

\citet{23} gives the star rotation period: $P = 5\fd162896$. We
obtained a close period from the analysis of the HD\,290665 light
curve obtained in the CDIPS survey basing on the TESS images:
$P_\mathrm{TESS}=5\fd176873$.

Ten observations of the star at the BTA show that the period is
indeed close to 5 days, but our measurements are in better
agreement with the period published~by \citet{23}. The
root-mean-square longitudinal magnetic fields of the star are as
follows:
 \begin{list}{$\bullet$}{
 \setlength\leftmargin{10mm} \setlength\topsep{2mm} \setlength\parsep{0mm} \setlength\itemsep{2mm}}
  \item $\langle B_{e} \rangle = 2871$~G, $\sigma = 121$, $\chi^2/n = 544.2$ (with the Babcock method),
  \item $\langle B_{e} \rangle = 2260$~G, $\sigma = 54$, $\chi^2/n = 2343.4$ (with the regression method).
 \end{list}

We determined the fundamental parameters of the star:
$T_\mathrm{eff} = 10400\pm350$~K, $\log g = 4.0\pm0.3$.

\section{DISCUSSION OF RESULTS}

Before comparing the magnetic properties of CP stars in the two
studied subgroups, 1a and 1b, let us once again recall the basic
data about the Orion\,OB1 association.

The OB1 association in Orion has an evident heterogeneous
structure. \citet{1964ARA&A...2..213B} divided the whole area of
the association into four subgroups. \citet{26} identified 814
stars in the association. The distribution of these stars by
subgroups of different ages is as follows:
\begin{list}{}{
\setlength\leftmargin{2mm} \setlength\topsep{1mm}
\setlength\parsep{0mm} \setlength\itemsep{1mm}} \item Orion\,OB1a
with an average age of 10~Myr contains 311~stars; \item
Orion\,OB1b, an age of 2~Myr, 139~stars; \item Orion\,OB1c, an age
of 5~Myr, 350~stars; \item Orion\,OB1d is a very small subgroup of
14 stars with an age smaller than 1~Myr.
\end{list}
Almost all of the listed objects are B and \mbox {A stars} of the
main sequence. The fraction of hot stars with effective
temperatures \mbox {$T_\mathrm{eff} > 10000$}~K is higher for
inner subgroups: 71.9\% for 1b and 92.9\% for 1d. For outer
subgroups~1a and 1c, this values are 51.1\% and 47.7\%
respectively \citep{1, 29}.

The procedure of selecting chemically peculiar stars in the
association was thoroughly described in~\citet{1}. We identified
85 chemically peculiar stars in it. In general, the proportion of
hotter stars among the chemically peculiar is greater than that
among the normal stars.

Now let us compare the magnetic field measurements for
subgroups~1a and 1b in the Orion association. After excluding the
non-peculiar star HD\,37149 from consideration, the number of CP
stars in both subgroups is equal: 15  stars in each group. For all
these stars we measured their magnetic fields in the same way. In
subgroup~1a, 7 out of 15 chemically peculiar stars (46.7\%)
appeared to be magnetic, while in subgroup~1b there were eleven
magnetic stars (73.3\%). As a criterion for the presence of a
magnetic field, we consider the value $\chi^2/n > 5$.

Thus, the fraction of magnetic stars among chemically peculiar
stars in younger subgroup~1b is 1.5~times higher than in
subgroup~1a. If we compare the proportions of stars with detected
magnetic fields relative to all stars in the corresponding
subgroup, the difference is even more striking: the proportion of
magnetic stars relative to all OBA stars in subgroup~1a is 2.25\%
(7 out of 311), and in subgroup~1b it is 7.91\% (11 out of 139),
or 3.5 times greater. This means that with an increase in the age
of stars from 2 to 10 Myr, a very sharp decrease in the proportion
of magnetic stars relative to the total sample is observed in the
Orion\,OB1 association.

Let us now consider the average magnetic fields of the stars in
the studied regions of the association. Obviously, non-magnetic
stars should be excluded from comparison, otherwise different
proportions of magnetic stars would distort the results. We
determine the average root-mean-square (rms) longitudinal magnetic
field $\langle B_{e}\rangle$({\rm all}) for the entire subgroup
similar to formula(\ref{for1}), taking the $\langle B_{e}\rangle$
value for each star as an individual measurement. The individual
rms fields of the stars in subgroups~1a and 1b are presented in
Tables~\ref{tab4} and \ref{tab5}. The magnetic field measurements
of the  stars from Table \ref{tab4} are taken from~\citet{5}.
There are only three stars in subgroup~1a (HD\,35298, HD\,35456,
and HD\,35502) the magnetic field of which are determined
absolutely reliably ($\chi^2/n > 30$). In subgroup~1b $\chi^2/n >
30$ for seven stars, and $\chi^2/n > 5$ for all the stars in the
table. In the case of HD\,36313 and HD\,37776 for the reasons
mentioned above in the comments for individual stars, the field
measurements were taken only by the H$\beta$ line (the stars are
marked with ``$^*$'' in Table~5).

\begin{table}[]
\caption{Root-mean-square magnetic fields $\langle B_{e}\rangle$
of the stars in subgroup~1a of the Orion\,OB1 association}
\label{tab4}
\medskip
\begin{tabular}{l|r@{$\,\pm\,$}l|c|r@{$\,\pm\,$}l|c}
\hline
\multicolumn{1}{c|}{\multirow{2}{*}{Star}} & \multicolumn{2}{c|}{$\langle B_e(z)\rangle \pm \sigma$,} & \multirow{2}{*}{$\chi^2/n$} & \multicolumn{2}{c|}{$\langle B_e(r)\rangle \pm \sigma$,} & \multirow{2}{*}{$\chi^2/n$} \\
                                             & \multicolumn{2}{c|}{G}                                  &                             & \multicolumn{2}{c|}{G}                                  &   \\
\hline
HD\,34859   & $1138$ & $692 $ &   3.8 & $ 302$ & $120$ &  9.9 \\
HD\,35008   & $1530$ & $1440$ &   3.8 & $ 258$ & $155$ &  7.0 \\
HD\,35177   & $1423$ & $1558$ &   4.3 & $ 940$ & $275$ & 12.4 \\
HD\,35298   & $4600$ & $563 $ & 120.5 & $2323$ & $330$ & 71.4 \\
HD\,35456   & $ 447$ & $96  $ &  34.3 & $ 440$ & $80 $ & 37.7 \\
HD\,35502   & $2221$ & $478 $ &  35.3 & $1647$ & $333$ & 41.7 \\
HD\,294046  & $2153$ & $1214$ &   4.6 & $1496$ & $164$ & 13.3 \\
\hline
\end{tabular}
\end{table}

\begin{table}[]
\caption{Root-mean-square magnetic fields $\langle B_{e}\rangle$
of the stars in subgroup~1b of the Orion\,OB1 association}
\label{tab5}
\medskip
\begin{tabular}{l|r@{$\,\pm\,$}l|c|r@{$\,\pm\,$}l|c}
\hline
\multicolumn{1}{c|}{\multirow{2}{*}{Star}} & \multicolumn{2}{c|}{$\langle B_e(z)\rangle \pm \sigma$,} & \multirow{2}{*}{$\chi^2/n$} & \multicolumn{2}{c|}{$\langle B_e(r)\rangle \pm \sigma$,} & \multirow{2}{*}{$\chi^2/n$} \\
                                             & \multicolumn{2}{c|}{G}                                  &                             & \multicolumn{2}{c|}{G}                                  &   \\
\hline
HD\,36313 * & $1338$ & $500 $   &  7.2 & \multicolumn{2}{c|}{}   &        \\
HD\,36485   & $2724$ & $261 $      &  113.8 & $2156$ & $132$        & 635.1  \\
HD\,36526   & $2801$ & $384 $      &   56.8 & $1695$ & $137$        & 539.3  \\
HD\,36668   & $1892$ & $451 $      &   37.6 & $ 953$ & $105$        & 203.5  \\
HD\,36955   & $ 843$ & $219 $      &   28.2 & $ 708$ & $90$         & 93.9   \\
HD\,37140   & $ 585$ & $207 $      &   14.0 & $ 270$ & $107$        & 9.6    \\
HD\,37333   & $ 870$ & $246 $      &   19.4 & $ 433$ & $120$        & 14.6   \\
HD\,37479   & $3040$ & $853 $      &   29.2 & $1307$ & $324$        & 46.6   \\
HD\,37633   & $ 520$ & $122 $      &   41.2 & $ 382$ & $97 $        & 25.5   \\
HD\,37776 * & $7285$ & $2686$      &   17.2 & $8644$ & $500$        & 298.9  \\
HD\,290665  & $2871$ & $121 $      &  544.2 & $2260$ & $54 $        & 2343.4 \\
\hline
\end{tabular}
\end{table}

The average rms magnetic field $\langle B_{e}\rangle({\rm all}$
for all the magnetic stars in subgroup~1a:
\begin{list}{$\bullet$}{
\setlength\leftmargin{10mm} \setlength\topsep{2mm}
\setlength\parsep{0mm} \setlength\itemsep{2mm}}
  \item $\langle B_{e} \rangle({\rm all}) = 2286$~G, $\sigma = 1000$, $\chi^2/n = 16.8$ (measured by the Babcock method),
  \item $\langle B_{e} \rangle({\rm all}) = 1286$~G, $\sigma = 229$, $\chi^2/n = 29.8$ (measured by the regression method).
 \end{list}

The almost double difference in the average rms fields measured by
different methods is explained by the presence of fast rotators
with complex spectral line profiles in the sample.

In the same way, we find the average rms magnetic field $\langle
B_{e}\rangle({\rm all})$ for all stars in subgroup~1b of the
association:
 \begin{list}{$\bullet$}{
 \setlength\leftmargin{10mm} \setlength\topsep{2mm} \setlength\parsep{0mm} \setlength\itemsep{2mm}}
   \item $\langle B_{e} \rangle({\rm all}) = 2911$~G, $\sigma = 893$, $\chi^2/n = 74.9$ (measured by the Babcock method),
    \item $\langle B_{e} \rangle({\rm all}) = 3014$~G, $\sigma = 211$, $\chi^2/n = 266.6$ (measured by the regression method).
  \end{list}

Thus, both methods of measuring magnetic fields give the same
result: the magnetic field of stars in younger subgroup~1b is much
stronger than that in subgroup~1a.

Earlier, we repeatedly showed that classical measurements by the
Babcock method for hot helium stars are extremely difficult
because of the small number of spectral lines suitable for
measurements and their complex profiles. The regression method
looks more preferable, but even in the cases of a strong field
with a complex structure, for example as in the star HD\,37776, it
can yield underestimated fields. Such cases require special
attention, so we consider the measurements made by the regression
method separately.

Comparing our measurements of magnetic fields of CP stars in
subgroups~1a and 1b of the Orion\,OB1 association, we come to the
conclusion that, on average, the field in the group of seven stars
with an age of about 2~Myr is 2.3~times higher than that in the
older group of eleven stars with an age of 10~Myr. We also see
that $\chi^2/n$, which characterizes reliability of magnetic field
detection, is an order of magnitude higher for stars of
subgroup~1b, this as well indirectly indicates that the magnetic
field of the stars in subgroup~1b is determined much more reliably
than that of the stars in subgroup~1a. It is possible that some
stars from our lists in subgroups~1a and 1b have weak fields and
we have not detected them, but this in no way affects the
conclusions we have obtained in this paper. Long ago, Babcock
noted that at the accuracy level in the order of 200~G, only every
fourth peculiar star has a magnetic field. With the improved
measurement accuracy, this number increased, but in every case it
does not exceed half of all measured CP stars. In the Orion\,OB1
association of young stars, in all its subgroups, the proportion
of CP stars with detected magnetic fields is higher than the
Babcock`s estimate; this fact also coincides with our conclusions
about higher occurrence of magnetic stars among the young
population.

\section{CONCLUSION}

Thus, a preliminary analysis of magnetic field measurements for
the stars in subgroups~1a and 1b of the Orion\,OB1 association
indicates that not only the proportion of peculiar stars relative
to normal ones decreases with age, which we showed earlier
in~\citet{1}, but the proportion of magnetic stars relative to all
peculiar stars in the subgroup also significantly drops, as well
as the strength of the magnetic field. The rate of the field
weakening in the time interval from 2 to 10~yr is unexpectedly
large.

On average, the temperature of stars in subgroup~1a appears to be
somewhat lower than that in subgroup~1b \citep{1, 29}. However,
the dependences that we have found cannot be explained by
temperature effects. Previously, a number of searches for a
dependence between the magnetic field and the temperature were
repeatedly carried out and, at best, a weak trend was seen,
indicating a decrease of the field with higher temperatures (see,
for example, \citet{28}). Based on the aforementioned, we believe
that the regularities we have found have evolutionary meaning.

It is possible that younger stars have a developed small-scale
field structure which rapidly decays with age, and its
contribution in the resulting field significantly decreases.
Observationally, this can appear as significant differences in the
field strengths obtained from the lines formed at different
heights in the atmosphere. This is a goal for future research. On
the whole, our result supports the idea of the fossil origin of
magnetic fields in \mbox {CP stars}. It is also obvious that the
theory of magnetic field formation in hot stars needs further
development. At this stage of the research, we have obtained the
data that can become an important quantitative observational test
for calibrating various mechanisms of formation and evolution of
large-scale stellar magnetic fields.

\section*{ACKNOWLEDGEMENTS}
The authors are grateful to the Russian Telescope Time Allocation
Committee for providing the observing time. The data were obtained
with polarization analyzers designed and developed by
G.~A.~Chountonov. We made use of the SIMBAD, VIZIER, and NASA/ADS
astronomical databases.

\section*{FUNDING}
Observations carried out with the 6-m BTA  are supported by the
Ministry of Science and Higher Education of the Russian Federation
(including agreement No.~05.619.21.0016, project~ID
RFMEFI61919X0016). The renovation of telescope equipment is
currently provided within the national ``Science'' project.
I.I.R., A.V.M., and I.A.Y. thank the Russian Foundation for Basic
Research for partial financial support of this work (RFBR grants
No.~\mbox{20-02-00233}, \mbox {18-29-21030}, \mbox{19-32-60007}).

\section*{CONFLICT OF INTEREST}
The authors declare no conflict of interest regarding the
publication of this paper.

%

\end{document}